\documentclass[letterpaper,twocolumn,10pt]{article}
\usepackage{usenix2019,epsfig,endnotes}

\usepackage{mathrsfs,amsmath}

\usepackage{graphicx,epsfig,sped,times}
\usepackage[utf8]{inputenc}
\usepackage{amsfonts}
\usepackage{booktabs}
\usepackage{amsmath}
\usepackage{balance}

\usepackage{algorithm}
\usepackage{units}
\usepackage{clrscode}
\usepackage[noeka]{mathrmletter}
\usepackage{gensymb}
\usepackage{ulem}
\usepackage{multirow}
\usepackage[noend]{algorithmic}
\usepackage{xspace}

\usepackage{url}

\usepackage[usenames, dvipsnames]{color}

\usepackage{subcaption}
\usepackage{xspace}

\usepackage[table]{xcolor}
\newcolumntype{C}[1]{>{\centering\let\newline\\\arraybackslash\hspace{0pt}}m{#1}}
\usepackage{ctable} 

\usepackage{pifont}
\usepackage{enumitem}
\newcommand{\squishlist}{\begin{itemize}[itemsep=1pt,parsep=2pt,topsep=3pt,partopsep=0pt,leftmargin=0em, itemindent=1em,labelwidth=1em,labelsep=0.5em]}
\newcommand{\squishend}{\end{itemize}}
\newcommand{\squishenum}{\begin{enumerate}[itemsep=1pt,parsep=2pt,topsep=3pt,partopsep=0pt,leftmargin=0em,listparindent=1.5em,labelwidth=1em,labelsep=0.5em]}
\newcommand{\squishsubenum}{\begin{enumerate}[itemsep=1pt,parsep=2pt,topsep=0pt,partopsep=0pt,leftmargin=0em,listparindent=1.5em,labelwidth=1em,labelsep=0.5em]}
\newcommand{\squishenumend}{\end{enumerate}}

\newcommand{\xref}[1]{\S\ref{#1}}
\newcommand{\tagname} {device\xspace}
\newcommand{\tagnames} {devices\xspace}
\newcommand{\tagName} {Device\xspace}

\newcommand{\name} {$NetScatter$\xspace}

\newcommand{\nameDesign} {NetScatter\xspace}

\newcommand{\supsym}[1]{\raisebox{6pt}{{\footnotesize #1}}}
\newfont{\coprimary}{phvr8t at 10pt}

\begin{document}
\date{}

\title{\Large \bf NetScatter: Enabling Large-Scale Backscatter Networks}


\author{%
Mehrdad Hessar\supsym{$\dagger$}, Ali Najafi\supsym{$\dagger$} and Shyamnath Gollakota\\
{University of Washington}\\
\coprimary{\supsym{$\dagger$}Co-primary Student Authors}\\
}


\maketitle

\thispagestyle{empty}

{\bf Abstract --} We present the first wireless protocol that scales to hundreds of concurrent transmissions from backscatter devices. Our key innovation is a distributed coding mechanism that works below the noise floor, operates on backscatter devices and can decode all the concurrent transmissions at the receiver using a single FFT operation. Our design addresses practical issues such as timing and frequency synchronization as well as the near-far problem. We deploy our design using a testbed of backscatter hardware and show that our protocol scales to concurrent transmissions from 256 devices using a bandwidth of only 500~kHz. Our results show throughput and latency improvements of {14--62x} and {15--67x}  over existing approaches and 1--2 orders of magnitude higher transmission concurrency.

\section{Introduction}
The last few years have seen  rapid innovations in low-power backscatter communication~\cite{abc,fsbackscatter,nsdi16,interscatter,wifibackscatter}, culminating in long range and reliable backscatter systems~\cite{lorabackscatter,plorasigcomm2018,lorea-europe}. These designs  enable wireless devices to communicate at microwatts of power and operate reliably at long ranges to provide whole-home or warehouse coverage. To achieve this, they employ  low-power coding techniques such as chirp spread spectrum, to decode weak backscatter signals below the noise floor~\cite{plorasigcomm2018,lorabackscatter} and deliver long ranges.

While these long range backscatter systems are promising for
enabling power harvesting devices (e.g., solar and vibrations)  as well as cheap and small Internet-connected  devices that operate on button-cells or flexible printed batteries, they primarily work at the link layer and are not designed to  scale with the number of devices --- all these prior designs~\cite{lorabackscatter,plorasigcomm2018,lorea-europe} are evaluated in a network of 1--2 devices. 


\begin{figure}[t]
	\centering
\vskip -0.25in
\includegraphics[width=0.95\linewidth]{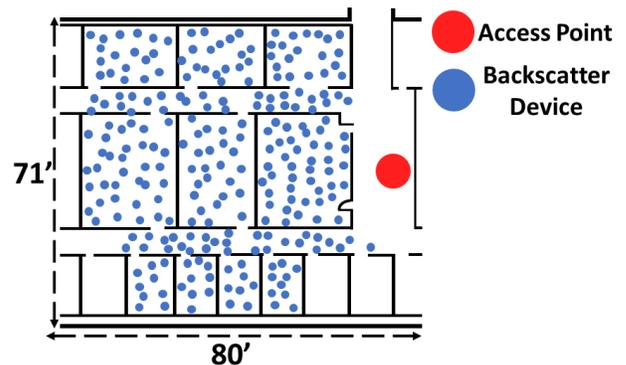}
	\vskip -0.1in
	\caption{{\bf Large-Scale Network Deployment of Backscatter Devices.} We deploy 256 backscatter devices across a floor of an office building covering multiple rooms.}
	\label{fig:floormap}
    \vskip -0.15in
\end{figure}

Our goal in this paper is to design a network protocol that enables these low-power backscatter networks to support hundreds to thousands of concurrent transmissions. This is challenging because the resulting design must operate reliably with weak backscatter signals that can be close to or below the noise floor.  To this end, we present \name,  the first wireless protocol that can scale to hundreds and thousands of concurrent transmissions from backscatter devices. Our design enables concurrent transmissions from  256 devices over a bandwidth of 500~kHz. Consequently, it can support transmissions from a thousand concurrent backscatter devices using a total  bandwidth of only 2~MHz.

Our key innovation is a distributed coding mechanism that satisfies four key constraints: i) it enables hundreds of devices to concurrently transmit on the same frequency band, ii) it can operate below the noise floor while achieving reasonable bitrates, iii) its coding operation can be performed by low-power  backscatter devices, and iv) it can decode all the transmissions at the receiver using a single FFT operation, thus minimizing the receiver complexity. 

\begin{figure}[t]
	\begin{subfigure}{\columnwidth}
    	\centering
    	\includegraphics[width=1\linewidth]{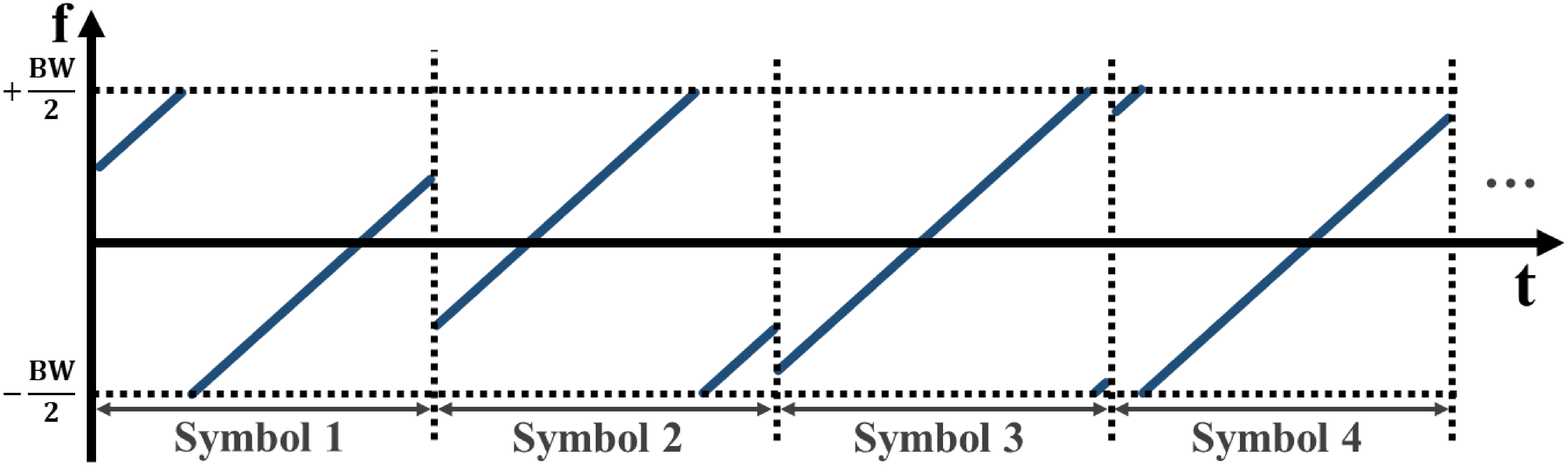}
		\label{fig:chirp_1}
        \vskip -0.15in
        \caption{Existing CSS Modulation (i.e., LoRa)}
	\end{subfigure}
    \begin{subfigure}{\columnwidth}
    	\centering
    	\includegraphics[width=1\linewidth]{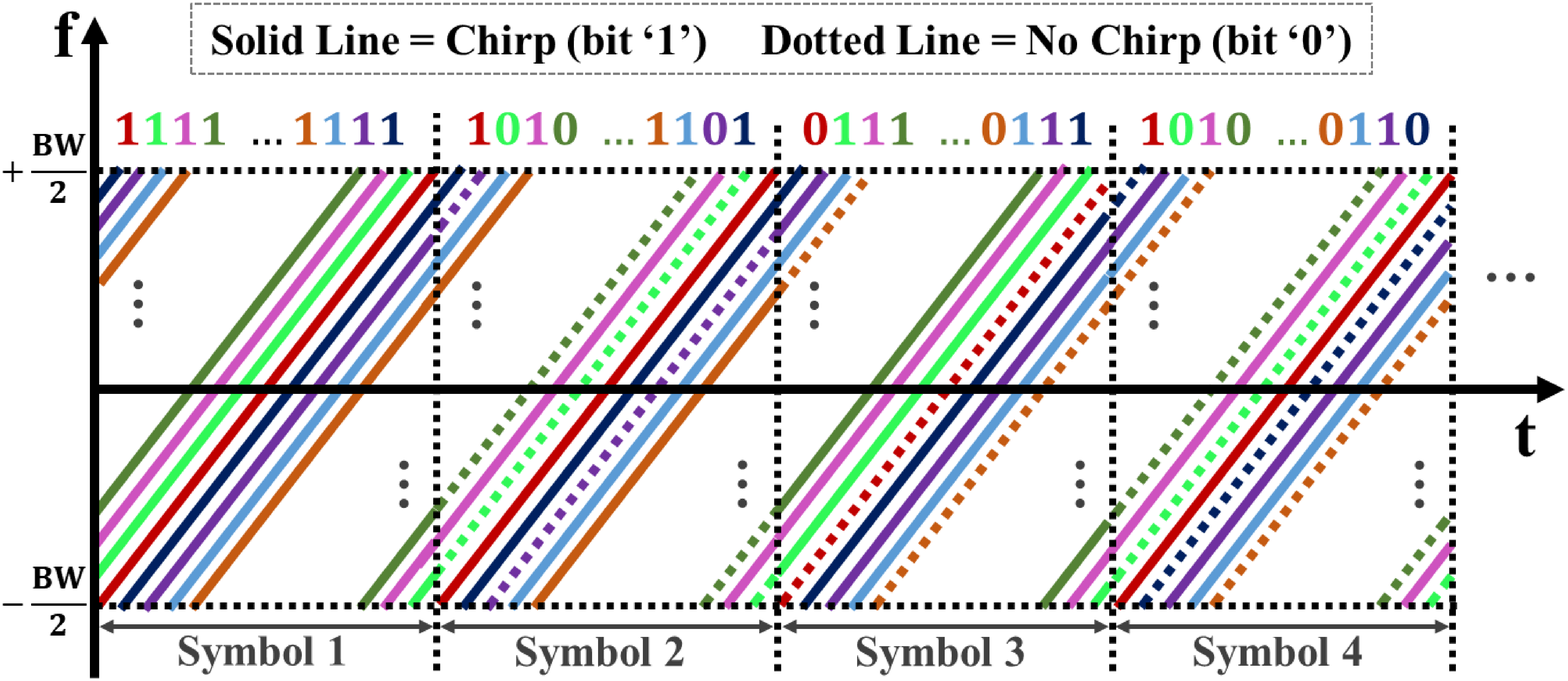}
		\label{fig:chirp_2}
        \vskip -0.15in
        \caption{Our Distributed CSS Coding}
	\end{subfigure}
   
	\vskip -0.1in
\caption{{\bf \nameDesign Overview.} In traditional CSS systems, a single device uses different cyclic shifts to convey bits. In distributed CSS coding,  each cyclic shift is assigned to a different backscatter device. Each device then uses the presence and absence of cyclic shift to send `1' and `0' bits.}
	\label{fig:modulation}
	\vskip -0.15in
\end{figure}

We introduce  
{\it distributed chirp spread spectrum coding}, which uses a combination of chirp spread spectrum (CSS) modulation and ON-OFF keying. In existing CSS systems (e.g., LoRa backscatter~\cite{lorabackscatter}), the AP transmits a continuous wave signal which each device backscatters and encodes bits using different cyclic shifts of a chirp signal. In contrast, in our distributed CSS coding, we assign a different cyclic shift of the chirp to each of the concurrent devices.  Each device then uses ON-OFF keying over these cyclic shifted chirps to convey bits, i.e., the presence and absence of the corresponding cyclic shifted chirp correspond to a `1' and `0' bit respectively, as shown in Fig.~\ref{fig:modulation}. Note that in comparison to existing CSS systems where each device transmits $log_2N$ bits using $N$ cyclic shifts, our distributed design enables $N$ concurrent devices, each of which  transmits a single bit using ON-OFF keying. Thus, our design transmits a total of $N$ bits within a  chirp duration, providing a theoretical gain of $\frac{N}{log_2{N}}$. 


Our design leverages the fact that creating concurrent cyclic-shifted chirps at a single \tagname requires distributing its transmit power amongst all the cyclic shifts, which reduces the ability of the receiver to decode each chirp. Instead we generate  concurrent cyclic-shifted chirps  across a distributed set of low-power \tagnames in the network. This allows us to efficiently leverage the coding gain provided by   chirp spread spectrum under the noise floor~\cite{chirppaper}. Further, we can decode all the concurrent transmissions using a single FFT operation, since cyclic shifting the chirps in the time domain translates to offsets in the frequency domain.  


Using the above distributed coding mechanism in practice, however, is challenging for two key reasons.
\squishlist

\item {\it Near-far problem.} A fundamental problem with enabling concurrent transmissions is that signals from a nearby backscatter device can overpower a farther concurrent device.  To address this issue, we introduce two main techniques. First, we present a power-aware cyclic shift allocation technique in~\xref{sec:nearfar}, where lower SNR \tagnames use much different cyclic shifts than higher SNR devices. We show that such an allocation can allow backscatter devices that have an SNR difference of up to 35~dB to be concurrently decoded. Second, to account for channel variations over time, we develop a zero-overhead power adaptation algorithm where backscatter devices use reciprocity to estimate their SNR at the AP, using the signal strength of the AP's query message. The backscatter devices then adjust their transmission power to fall within the tolerable SNR difference. Since this calibration is done independently at each backscatter device using the AP's query, it does not require additional communication overhead at the AP.

\item {\it Timing synchronization.} The above design requires all the \tagnames to start transmitting at the same time so as to enable concurrent decoding. However, hardware variation and propagation delays of different \tagnames can make it challenging for hundreds of \tagnames to be tightly synchronized in time. To avoid this coordination overhead, we leave gaps between cyclic shifts to ensure that concurrent \tagnames are sufficiently distinguishable and can be decoded. We explore the trade-off between the required gaps and the chirp bandwidth in~\xref{sec:timing}.
\squishend

We implement {\name} on a testbed of backscatter devices. We create backscatter  hardware that implements {\name} and includes circuits to perform automatic power adaptation before each transmission. We deploy our backscatter testbed with 256 devices in an office building spanning multiple rooms as shown in Fig.~\ref{fig:floormap}. We implement our receiver algorithm {using USRP X-300 software-defined radios}. Our results reveal that over a 256 node backscatter deployment, {\name} achieves a {14--62x} gain over prior long-range backscatter systems~\cite{lorabackscatter} for its end-to-end link layer data rates. The key benefit however is in the network latency which sees a reduction of {15--67x}.

\vskip 0.05in\noindent{\bf Contributions.} Our paper demonstrates, to the best of our knowledge, the first network protocol that achieves orders of magnitude more concurrent transmissions than existing backscatter systems. The closest work to our design is Choir~\cite{lorasigcomm17} {\it in the radio domain}, which decodes concurrent transmissions from 5--10 LoRa radios at a software radio. Choir leverages frequency imperfections to disambiguate between LoRa radios. However, backscatter devices achieve low power operations by running at a  lower frequency (1-10~MHz) than radios (900~MHz) and thus have much smaller frequency differences between backscatter devices. This severely limits the ability to  rely on frequency imperfections to disambiguate between a large number of backscatter devices (see~\xref{sec:choir}). In contrast, our distributed chirp spread spectrum coding mechanism provides a systematic approach to enable large scale backscatter networks.


\begin{figure}[t]
\vskip -0.15in
	\begin{subfigure}{0.325\columnwidth}
    	\centering
    	\includegraphics[width=1\linewidth]{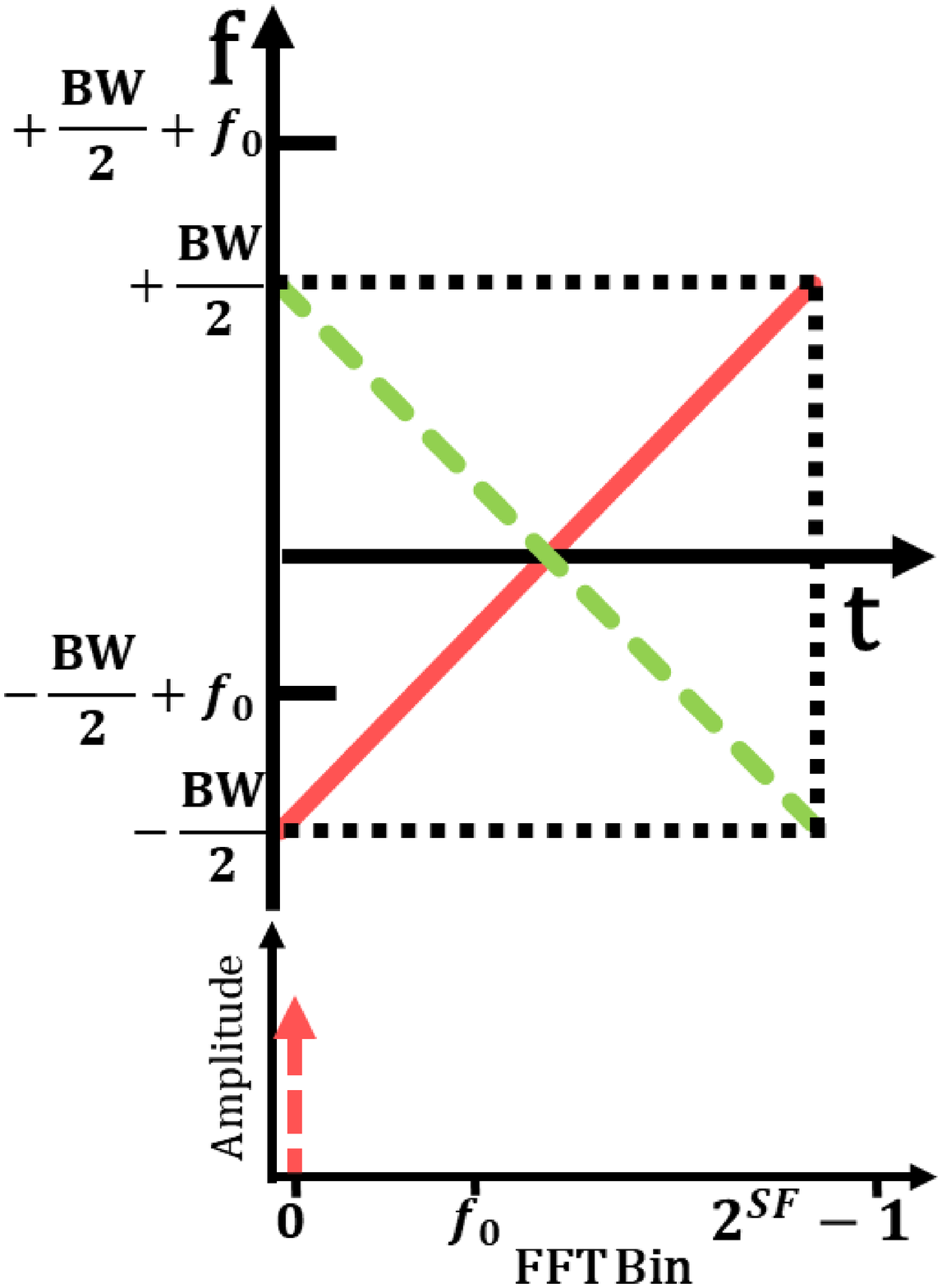}
		\label{fig:chirp_1}
        \vskip -0.2in
        \caption{}
	\end{subfigure}
    \begin{subfigure}{0.325\columnwidth}
    	\centering
    	\includegraphics[width=1\linewidth]{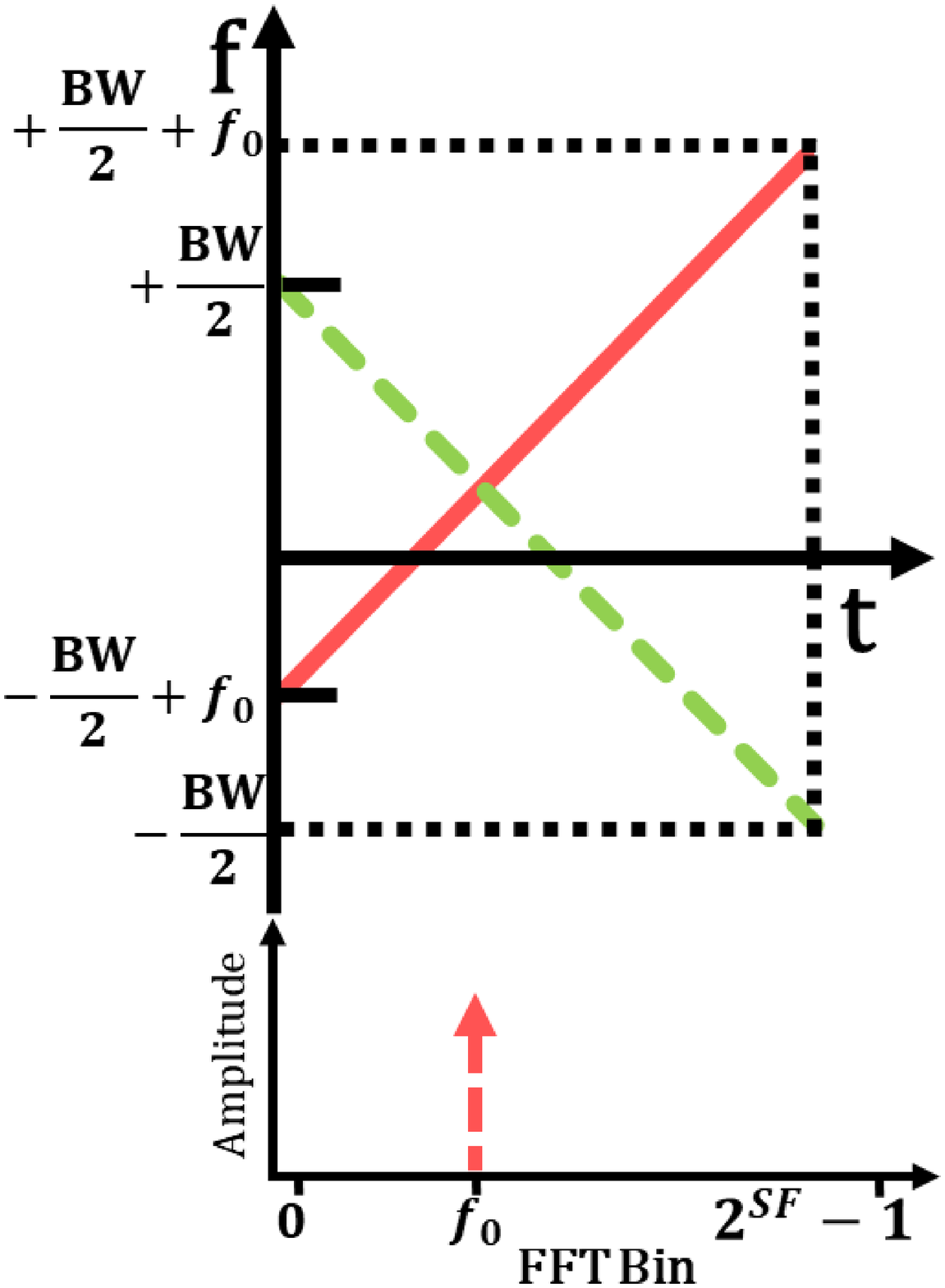}
		\label{fig:chirp_2}
        \vskip -0.2in
        \caption{}
	\end{subfigure}
    \begin{subfigure}{0.325\columnwidth}
    	\centering
    	\includegraphics[width=1\linewidth]{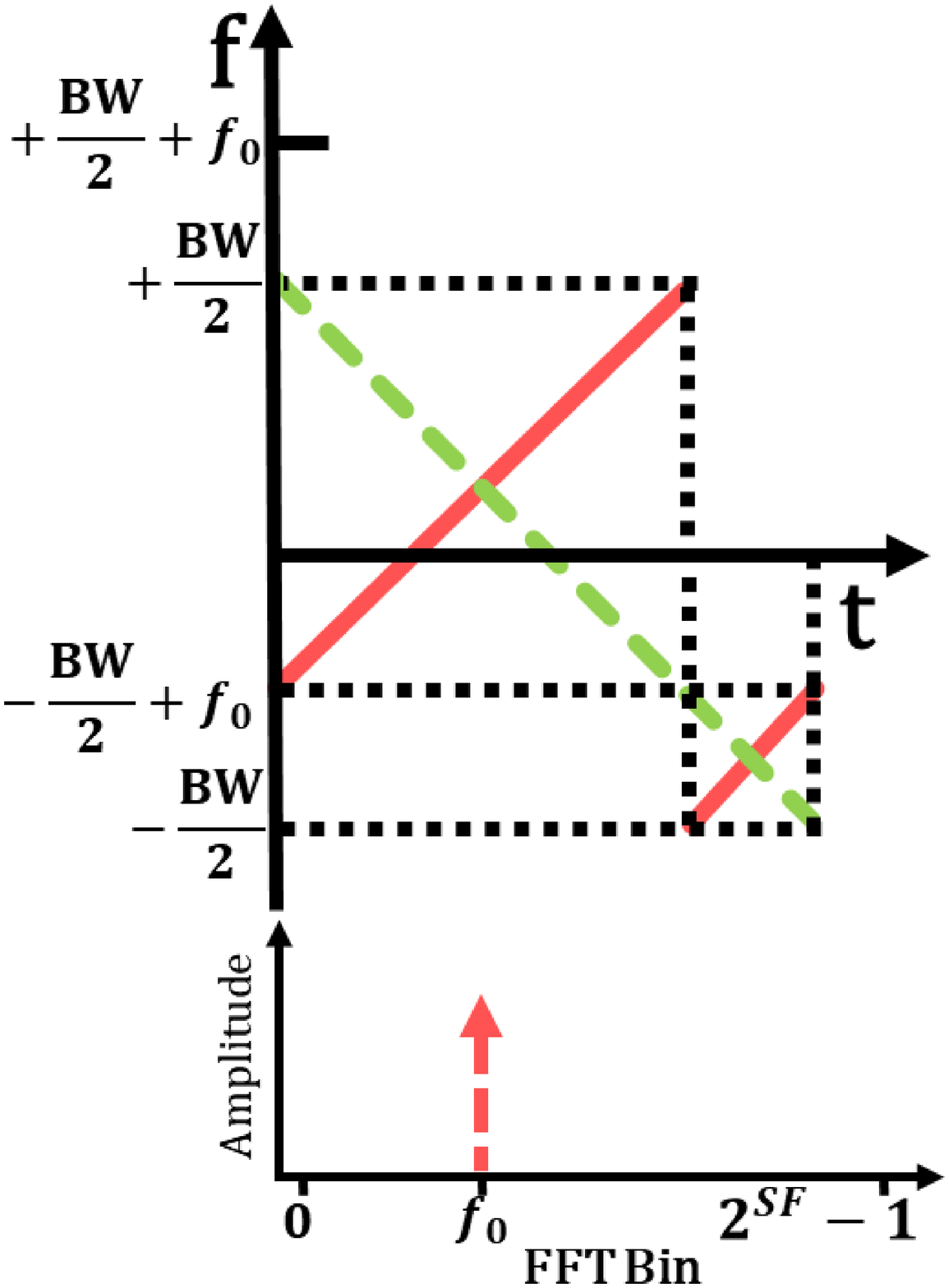}
		\label{fig:chirp_3}
        \vskip -0.2in
		\caption{}
	\end{subfigure}
\vskip -0.1in
\caption{{\bf CSS Primer.} We show upchirp and downchirp symbols and FFT results of their multiplication. (a) Baseline upchirp symbol, (b) frequency shifted upchirp symbol and (c) cyclically shifted upchirp symbol.}
\label{fig:chirp_fft}
\vskip -0.15in
\end{figure}

\section{CSS Primer \& Existing Approaches}\label{sec:primer and exist}
\subsection{Primer on Chirp Spread Spectrum}\label{css_primer}
In CSS, data is modulated using linearly increasing frequency signals or upchirps. The receiver demodulates these symbols in a two step process. First, it de-spreads these upchirp symbols by multiplying them by a downchirp and it then performs an FFT on the de-spread signal. Since the slope of the downchirp is the inverse of the slope of the upchirp,  multiplication results in a constant frequency signal, as shown in Fig.~\ref{fig:chirp_fft}(a). Thus, taking an FFT on this will lead to a peak in an associated FFT bin. Changing the initial frequency of an upchirp will result in a change in the demodulated signal's FFT bin peak index which corresponds to the initial change in frequency, as shown in Fig.~\ref{fig:chirp_fft}(b). This property is used to convey information. When the sampling rate is equal to chirp bandwidth (BW), frequencies higher than $\frac{BW}{2}$ will alias down to $\frac{-BW}{2}$ as shown in Fig.~\ref{fig:chirp_fft}(c). This means cyclically shifting in time is equivalent to changing the initial frequency and thus to      conserve bandwidth, CSS uses cyclic shifts of the chirp in the time-domain instead of frequency shifts. This means that to modulate the data we just need to cyclically shift the baseline upchirp in time. Note that one can transmit multiple bits within each upchirp symbol. In particular, say the receiver performs an $N$ point FFT. It can distinguish between $N$ different cyclic shifts each of which corresponds to a peak in one of the $N$ FFT bins. Thus, we can transmit $SF=log_{2}N$ bits within each upchirp symbol, where SF is called the spreading factor.

Based on above explanations, CSS can be characterized by two parameters: chirp bandwidth/sampling rate and spreading factor. Thus, each chirp symbol duration is equal to $\frac{2^{SF}}{BW}$ and the symbol rate is $\frac{BW}{2^{SF}}$. Since CSS sends $SF$ bits per symbol, the bitrate is equal to $\frac{BW}{2^{SF}}SF$. This means increasing $SF$ or decreasing $BW$ decreases the bitrate. Further, the sensitivity of the system depends on the symbol chirp duration and increases with $SF$ and decreases with $BW$.

\subsection{Existing Collision Approaches}\label{sec:existing}
While existing CSS-based backscatter systems do not support collision decoding, we outline potential approaches to deal with collisions in CSS {\it radio} systems, i.e. LoRa, and explore whether they can be adopted for backscatter.



\vskip 0.05in\noindent{\bf Using different spreading factors.}
One way to enable concurrent transmissions is to assign different spreading factors to each \tagname. There are three problems with using multiple spreading factors in the same network: {i) the receiver needs to use multiple FFTs and downchirps with different spreading factors to despread upchirp symbols of different \tagnames, which increases the receiver complexity with the number of concurrent transmissions, ii) in LoRa, different $BW$ and $SF$ can be concurrently decoded without sensitivity degradation, only if the chirp slope is different~\cite{sornin2017signal}. Specifically, if two chirp symbols transmitted concurrently with different $BW$ and different $SF$, which result in the same chirp slope, $\frac{BW^{2}}{SF}$ (shown in Fig.~\ref{fig:time_offset} as well), the receiver cannot decode their concurrent transmissions. This results in only 19 different $BW$ and $SF$ pairs that could be used concurrently, iii) further, requiring receiver sensitivity better than -123~dBm and bit rates of at least 1~kbps limits these concurrent configurations to only 8, which does not support hundreds of concurrent devices on a 500~kHz band. 
{Note that ignoring the receiver complexity, this approach is orthogonal to our design since we could in principle run multiple concurrent \nameDesign networks with the above 8 $SF$ and $BW$ pairs. Evaluating this is not in the scope of this paper.

\begin{figure}[!t]
   \centering
   \includegraphics[width=0.9\linewidth]{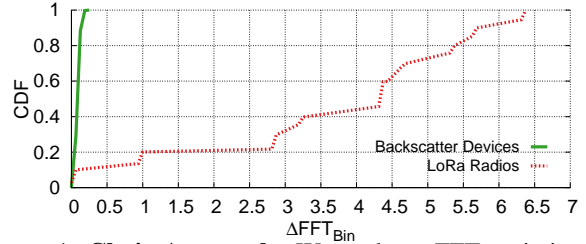}
   \vskip -0.15in
	\caption{{\bf Choir Approach.} We evaluate FFT variation of chirp symbols when $BW=500~kHz$ and $SF=9$ for both active LoRa radios and backscatter devices.} 
	\label{fig:choir_compare}
    \vskip -0.2in
\end{figure}

\vskip 0.05in\noindent{\bf Choir~\cite{lorasigcomm17}.}\label{sec:choir} Recent work on decoding concurrent LoRa transmissions leverages the hardware imperfections in radios to disambiguate between multiple transmissions. Specifically, radios have slight variations which result in timing and frequency offsets, which translate to fractional shifts in the FFT indexes. Choir~\cite{lorasigcomm17} uses these fractional shifts, with a resolution of one-tenth of an FFT bin, to map the bits to each transmitter. However, as demonstrated in~\cite{lorasigcomm17}, in practice this approach does not scale to more than 5 to 10 concurrent \tagnames. To understand this limitation in theory, consider N concurrent \tagnames. The probability that each of these transmitters has a different FFT peak index fraction, given the resolution of one-tenth of an FFT bin, is equal to $\frac{10!}{(10-N)!10^N}$. When $N$ is 5 this probability is only 30\%. Moreover, if any two transmitters use the same cyclic shifted upchirp symbol at the same time, it will result in a collision that cannot be decoded. In the case of LoRa modulation, if there are $N$ transmitters and assuming each \tagname transmits a random set of bits during each symbol interval, the probability of two transmitters using the same cyclic shift is equal to:
$1-\prod_{i=1}^{N}(1-\frac{i-1}{2^{SF}})$ which is approximately $\frac{N(N-1)}{2^{SF+1}}$.

For $SF=9$ and $N=10$, this probability is around 9\%. This means that there is around 9\% probability that within each CSS symbol, two transmitters will use the same upchirp cyclic shift, which the receiver cannot disambiguate. This probability increases to 32\% with 20 \tagnames, preventing concurrent decoding of a large number of transmitters.\par


\begin{figure}[t]
	\centering
	\includegraphics[width=1\linewidth]{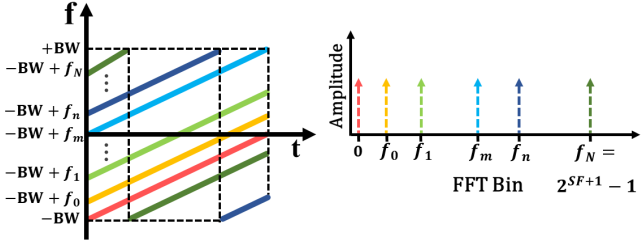}
	\vskip -0.1in
	\caption{{\bf Bandwidth Aggregation.} Here we use an aggregate bandwidth of 2BW but each device transmits only using BW. Upchirps with different cyclic shifts shown in different colors. Each upchirp is assigned to a device.}
	\label{fig:bw}
    \vskip -0.15in
\end{figure}

{Moreover, Choir is based on oscillator imperfection causing frequency variation on different \tagnames, and Choir cannot differentiate two concurrent transmissions if both transmissions fall into same FFT bin fraction. Choir uses an active radio system which generates frequencies in 900~MHz band. However, since backscatter systems are designed to consume less power and only generate baseband signals, their output frequency is less than 10~MHz. Now, in the ideal scenario where the same crystal oscillator is used for both radios and backscatter devices, the frequency variation of the backscatter devices is 90 times smaller than radios and can be even less than 1 FFT bin depending on the $SF$ and $BW$. This means a backscatter network cannot use all 10 different FFT bin fractions that Choir have used. 
Fig.~\ref{fig:choir_compare} shows CDF of FFT bin variation for our actual backscatter hardware which are recorded over time. This results show that FFT variation is always less than a third of an FFT bin. Thus, Choir cannot enable large concurrent transmissions with backscatter.}

In conclusion, the desired solution must satisfy three  constraints: 1) ability to differentiate between FFT peaks corresponding to different backscatter devices, 2) ability to associate the FFT peaks to the corresponding devices, and 3) ensure that two devices do not use the same FFT peak at the same time. \nameDesign design satisfies all these constraints. 
 
\section{\nameDesign Design}\label{sec:design}

\subsection{Distributed CSS Coding}
Our approach is to take advantage of low-power and high sensitivity of CSS modulation to design a communication and networking system that enables hundreds of backscatter devices to transmit at the same time.

At a high level, we use a combination of CSS modulation and ON-OFF keying to enable concurrent transmissions. Our intuition is as follows: if we look at the FFT plots of Fig.~\ref{fig:chirp_fft}, all the FFT bins except one bin are empty; however these empty bins could be utilized for orthogonal transmissions. While it is difficult to design low-power backscatter devices that can transmit multiple cyclic shifts at the same time, we can leverage all these empty bins by having different devices transmit different shifts and make use of the unused FFT bins. In particular, each device is assigned to a particular cyclic shifted upchirp symbol. It sends data by either sending the upchirp symbol or not sending it, i.e., by using ON-OFF keying of its assigned cyclic shifted chirp. Since, there are $2^{SF}$ FFT bins, ideally we can support $2^{SF}$ concurrent transmissions. This modulation will satisfy the above three requirements. The peaks can be differentiated and assigned to their corresponding devices. Moreover, none of them will use the same FFT bin at the same time.

We note the following about our distributed  design.
\squishlist
\item {\bf Receiver complexity.} The received signal is composed of multiple transmissions. They can be demodulated by despreading with a baseline downchirp multiplication and performing an FFT operation. Then, we can determine the presence and absence of a peak in each FFT bin and find if the corresponding backscatter device is sending `0' or `1'. The key point  is that the process of despreading and performing FFT, which are the major contributors of the demodulation process and  provide a coding gain for each of the backscatter devices enabling them to operate below the noise floor, are being done once and do not depend on the number of concurrent transmissions. This means that the receiver complexity is nearly constant with the number of devices.

\item {\bf Throughput gain.} In our approach, ideally there can be as many as $2^{SF}$ transmissions at each symbol period. Since each backscatter device uses ON-OFF keying over a symbol, their individual data rate  is $\frac{BW}{2^{SF}}$. Thus, the aggregate network throughput  is equal to $BW$. In comparison, LoRa have a throughput of $\frac{BW}{2^{SF}}SF$. Thus, we can achieve a throughput gain of $\frac{2^{SF}}{SF}$, which shows that the gain exponentially increases with the SF value used in the system. This is expected since the number of concurrent \tagnames we can support is an exponential function of SF, i.e., $2^{SF}$.


\begin{figure}[t]
    \centering
    \includegraphics[width=1\linewidth]{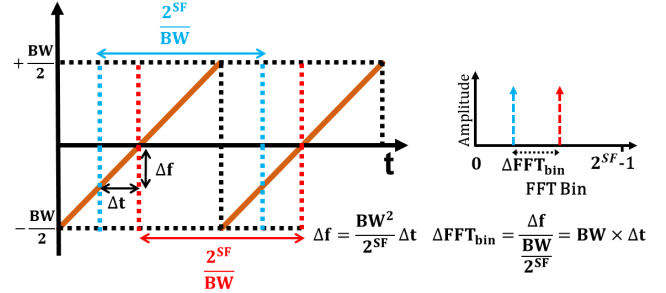}
	\vskip -0.1in
	\caption{{\bf Timing Mismatch,} in detecting beginning of a chirp symbol and its translation to FFT bin variation.}
    \label{fig:time_offset}
    \vskip -0.2in
\end{figure}

\item {\bf NetScatter and CDMA.} Our distributed CSS coding can be thought of as code-division multiplexing mechanism that is low-power and where each of the $2^{SF}$ cyclic shifts is in an orthogonal set of codes in a CDMA system. These orthogonal codes are then assigned to $2^{SF}$ different backscatter devices which enables $2^{SF}$ concurrent transmissions.

{\item{\bf Gain in the context of Shannon capacity.} A key gain we are achieving in our design stems from using the power across all the concurrent backscatter devices. Specifically we note that the Shannon capacity of a multi-user network that operates under the noise floor linearly increases with the number of devices. Said differently, the multi-user capacity of an access point network is given as~\cite{davidtse}, $C=BW\log_{2}(1+\frac{N P_{S}}{P_{N}})$. Here $BW$ is the channel bandwidth, $P_N$ and $P_S$ are the noise and signal power and $N$ is the number of concurrent \tagnames. At SNRs below the noise floor, the above equation can be approximated as $\frac{BW}{ln(2)}\frac{NP_S}{P_N}$, since $ln(1+x)\approx x$ when $x$ is small. This means that for systems that operate below the noise-floor, the network capacity scales linearly with the number of users. This linear increase stems from the fact that the $N$ backscatter devices put in $N$ times more power back to the AP than a single device.}

\item {\bf Bandwidth Aggregation.}
The bitrate achieved by each backscatter device in our distributed design is given by $\frac{BW}{2^{SF}}$ and the number of concurrent devices is $2^{SF}$. Thus, while we can increase the number of devices by increasing SF, it would decrease the bitrate of each device. Thus, to increase both the bitrate and the number of device we should increase the bandwidth, $BW$. Say, we want to support twice the number of devices while maintaining the same bitrate by using twice the bandwidth. This can be achieved in two ways. First, we can  use two filters and independently operate two sets of devices across the two bands. This approach requires two different FFTs to be performed independently across the bands. The second approach is to use one aggregate band with twice the bandwidth, $2BW$, but use the same $SF$ and chirp $BW$ as before and alias down to $-BW$ whenever the chirp frequency hits the maximum as shown in Fig.~\ref{fig:bw}. To demodulate this signal, we just need to multiply the signal which is composed of the aggregate band by the downchirp and perform $2\times2^{SF}$ FFT operation once. The complexity of this method is lower than the former since there is no need to use filters and separate the bands. 
\squishend

\subsection{Addressing Practical Issues}\label{issue}


\subsubsection{Timing Mismatch}
\label{sec:timing}
\par
The above design requires all the backscatter devices to be time synchronized. To understand why, consider two consecutive upchirps being sent by a device, as shown in Fig.~\ref{fig:time_offset}. Now say that we demodulate the signal in these two timing durations, shown in blue and red, we will get different FFT peak locations. Specifically, with a $\Delta t$ time difference between these durations, the corresponding FFT bin peak location would change by, $\Delta FFT_{bin}=\Delta t BW$. When this change is greater than a single FFT bin, backscatter devices that are assigned to consecutive cyclic shifts interfere with each other and hence cannot be decoded.  Thus, all the devices should be time synchronized. In our design the access point sends a query message telling devices to transmit concurrently. The devices use this query to synchronize and respond concurrently. First, we explain the sources of time delay in our system and then we explain our solution. There are multiple factors that can contribute to time delays introduced in practice and can be different for different backscatter devices.
\squishlist

\item {\bf Hardware delay.} Unlike Wi-Fi devices which use much higher clock frequencies for processors, backscatter devices use low-power microcontrollers (MCU) that can introduce a variable delay into the system. For backscatter \tagnames, the source of these hardware delay variations come from the time the envelope detector receives the query message from the access point, communicates it to the MCU and then the \tagname backscatters the chirp. As we show in~\xref{sec:tofo_eval}, this hardware delay variations can be as high as 3.5~$\mu$s, which can translate to more than one FFT bin at 500~kHz bandwidth.

\item {\bf Propagation delay and multipath.} {Since backscatter devices can be at different distances to the access point, their time of flight (TOF) can be different. However, since our target application is  for whole-home or whole-office sensing, the propagation distance is less than 100~m which translates to a $ToF<666ns=\frac{2\times100}{3\times 10^{8}}$ and corresponds to only a $0.33$ FFT bin change, assuming a bandwidth of 500~kHz. }
 {The multipath delay spread for indoor environments is between 50 to 300~ns~\cite{saleh1987statistical, devasirvatham1984time}. For 500~kHz, this delay spread translates to less than 0.15~FFT bin change, which is negligible. }

\squishend

\vskip 0.05in\noindent{\bf Our solution: Bandwidth-based cyclic-shift assignment.} Hardware delay variations over time are hard to correct for. As described above, by nature of operating on MCUs and other low-power computational platforms, these devices have a hardware delay variation over time that changes between packets. Our solution to this problem is to put a few empty FFT bins adjacent to each FFT bin assigned to a device. That is, if FFT bin $i$ is assigned to a device, the adjacent $SKIP-1$ FFT bins are empty and not assigned to any device. This can be done by using only every $SKIP^{th}$ cyclic shift of the chirp. This ensures that the hardware delay does not result in adjacent devices interfering with each other. 

Achieving such an assignment requires us to answer the following key question: how do we pick the value $SKIP$? As described earlier, given the hardware delay variation $\Delta t$, the shift in the number of FFT bins is $\Delta t BW$. This means that there is a trade-off in our system regarding the total network throughput, bitrate for each \tagname and sensitivity. In particular, increasing $BW$ increases the number of FFT bins that have to be left empty and decreases the total network throughput. On the other hand, decreasing $BW$ reduces the number of FFT bins but decreases the bitrate per \tagname with the same $SF$. To compensate for the decreased \tagname's bitrate, we can decrease the $SF$. Note that, we can choose total bandwidth, chirp $BW$ and $SF$ of the system by considering the hardware delay variations, required bitrate per \tagname, sensitivity for each \tagname and total number of devices. For our implementation, we pick the same total bandwidth and chirp $BW$ of $500$~kHz and $SF=9$ which supports around 1~kbps (976~bps) bitrate at the \tagnames while ensuring that the number of empty bins between \tagnames, $SKIP$, is two.

{\footnotesize
\begin{table}
{
\centering
\footnotesize
\caption{{\bf NetScatter Different Modulation Configurations,} with maximum time/freq. mismatch that can be tolerated.}
\vskip -0.1in
\begin{tabular}{C{0.7cm}|C{0.3cm}|C{1.2cm}|C{1.3cm}|C{1.3cm}|C{1.3cm}} 
\hline
\rowcolor{lightgray}
{\bf\color{black} BW [kHz]} & {\bf\color{black} SF} & {\bf\color{black} Time Variation} &{\bf\color{black} Frequency Variation} &{\bf\color{black} Bit Rate [bps]}&{\bf\color{black} Sensitivity [dBm]}\\
\rowcolor{white}
500& 9& 2~$\mu s$& 976~Hz & 976& {-123}\\
\hline
\rowcolor{white}
500& 8& 2~$\mu s$& 1953~Hz & 1953& {-120}\\
\hline
\rowcolor{white}
250& 8& 4~$\mu s$& 976~Hz & 976& {-123}\\
\hline
\rowcolor{white}
250& 7& 4~$\mu s$& 1953~Hz& 1953& {-120}\\
\hline
\rowcolor{white}
125& 7& 8~$\mu s$& 976~Hz& 976& {-123}\\
\hline
\rowcolor{white}
125& 6& 8~$\mu s$& 1953~Hz& 1953& {-118}\\
\hline
\end{tabular}

\label{tab:timing and freq tol.}
}
\vskip -0.15in
\end{table}
}

\begin{figure}[t!]
	\begin{subfigure}{0.5\columnwidth}
    	\centering
    	\includegraphics[width=1\linewidth, height=2.1cm]{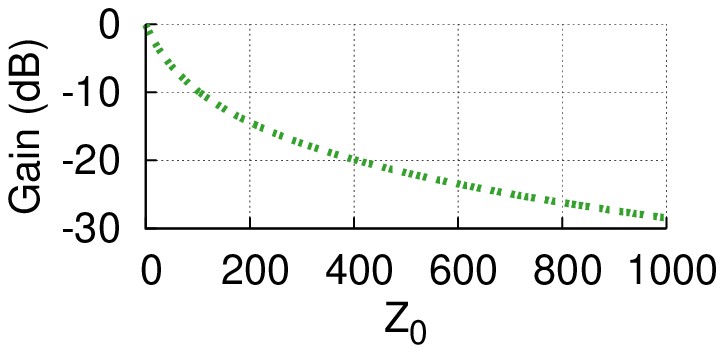}
        \caption{}
		\label{fig:impedance_plot}
	\end{subfigure}
    \begin{subfigure}{0.35\columnwidth}
    	\centering
    	\includegraphics[width=1.15\linewidth]{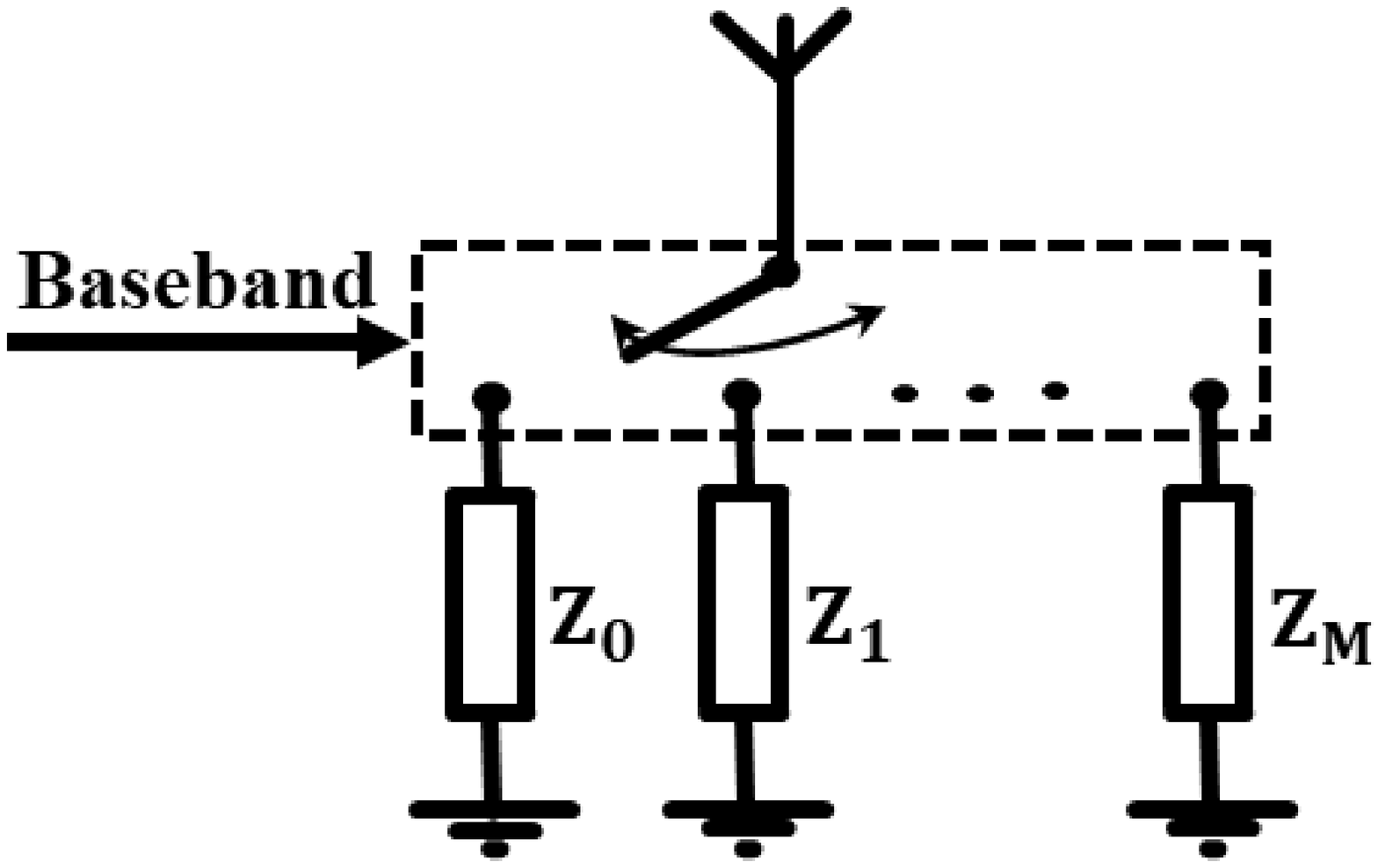}
        \caption{}
		\label{fig:impedence}
	\end{subfigure}
\vskip -0.15in
\caption{{\bf Power Adjustment for Backscatter.} (a) gain normalized to maximum power as a function of Z$_{0}$ impedance and (b)  switch network to support multiple power levels.}
\label{fig:gain_adjust}
\vskip -0.15in
\end{figure}

\subsubsection{Frequency Mismatch} 
The \tagnames experience frequency offsets because of hardware variations in the crystals used in their oscillators. As explained in~\xref{css_primer}, change in frequency translates to FFT bin change of the demodulated \tagname packet. This again, causes one \tagname to be misinterpreted as other \tagname. Considering a bandwidth of $BW$ and spreading factor of $SF$, the frequency difference between FFT bins is equal to $\frac{BW}{2^{SF}}$. This means that a $\Delta f$ frequency offset results in a change in the FFT bin of $\Delta FFT_{bin}=\frac{2^{SF}\Delta f}{BW}$. Therefore either increasing the spreading factor $SF$ or decreasing the $BW$ can increase the shift in the FFT bin. Crystals' frequency tolerance can be as high as $100~ppm$~\cite{crystalPPM}. Since backscatter \tagnames run at a few MHz frequencies, this frequency variation translates to less than one FFT bin for the bandwidths and spreading factors in this paper which makes it negligible for our backscatter network.

{Table~\ref{tab:timing and freq tol.} shows the timing and frequency mismatch that can be tolerated for different modulation configurations. As can be seen, there are multiple options for achieving the same bitrate and sensitivity. These options will result in different tolerable timing and frequency mismatch, requiring a different $SKIP$ value; this is validated using experiments in~\xref{sec:tofo_eval}.}


\begin{figure}[t]
    \centering
    \includegraphics[width=1\columnwidth]{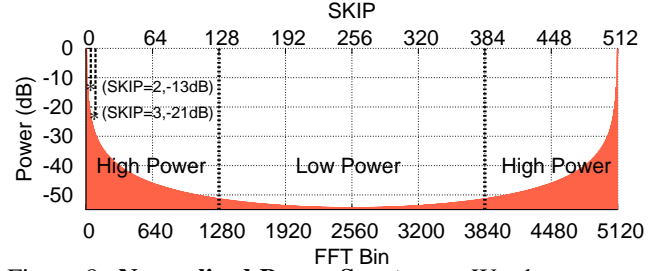}
    \vskip -0.15in
	\caption{{\bf Normalized Power Spectrum.} We show power spectrum of an upchirp multiplied by a baseline downchirp in FFT domain. This plot shows the main lobe and side lobes of a single chirp transmission. We assign \tagnames to high and low power regions based on their power level.} 
	\label{fig:fft_lobe}
    \vskip -0.15in
\end{figure}

\subsubsection{Near-Far Problem}
\label{sec:nearfar}
Since our networks are designed to work in below-noise conditions, we need to address the near-far problem in our decoding process at the receiver. Specifically, to account for the residual timing and frequency offsets, a CSS receiver has to achieve a sub-FFT bin resolution. To do so without increasing the sampling rate, the receiver uses zero-padding which adds zeros at the end of the time domain samples of the single chirp~\cite{lorasigcomm17}.  Zero-padding operation in the time domain is effectively a multiplication operation with a pulse which translates to convolution with a sinc function in the FFT domain. This makes it easier to locate the FFT peak location. However, convolving with a sinc function introduces side lobes as shown in Fig.~\ref{fig:fft_lobe}. Assume that there are two \tagnames  with  cyclic shifts $C_{1}=0$ and $C_{2}$. If the power of $C_{2}$ is lower than power of $C_{1}$'s side lobes, it cannot be decoded.

\vskip 0.05in\noindent{\bf Our solution.} To address this issue, we propose two techniques that work together to increase our dynamic range.

{\it Coarse-grained power-aware cyclic shift assignment.} Our intuition here is as follows: Fig.~\ref{fig:fft_lobe} suggests that we should assign adjacent FFT bins to \tagnames that have a small SNR difference. {In particular, when SKIP is 2, the neighboring backscatter device will be drowned by the power of the higher SNR device if its power is lower than 13.5~dB of that of the high SNR device.} Further, it shows that the side-lobe power of a high SNR \tagname decreases as we go to farther FFT bins. Thus, we need to ensure that a lower SNR \tagname has to correspond to FFT bins that are farther from the FFT bins corresponding to higher SNR \tagnames. This ensures that the side-lobes of the high-SNR \tagname do not affect the decoding of the low-SNR \tagnames. Specifically, we assign different cyclic shifts to different \tagnames at association phase to ensure that the FFT bins corresponding to the lower-SNR \tagnames are close to each other and are far  from higher-SNR \tagnames. To do this, the AP computes the signal strength of the incoming \tagname in the association phase (see~\xref{sec:assoc}) and assigns its cyclic shift based on its signal strength and also the strengths of the \tagnames already in the network.

We run simulations to understand the benefits of this allocation. Specifically, we assign two \tagnames to FFT bins 2 and 258, with $SF=9$ and $BW=500~kHz$. To be realistic, we added Gaussian frequency mismatch with variance of 300~Hz to each \tagname to account for timing and frequency mismatches between them. We change the power of the second \tagname and measure the bit error rate (BER) for the first \tagname. {Fig.~\ref{fig:snr_power_diff} shows the BER over 10$^{4}$ symbols, for different power differences between the two devices. As can be seen, the BER remains unaffected even when the second \tagname is around 40~dB stronger than the first \tagname. This shows that our power-aware allocation can in theory tolerate power difference of 40~dB between \tagnames. In practice however this is a little lower at 35~dB (see~\xref{nearfar_eval}).

{\it Fine-grained self-aware power-adjustment.} While the above assignment is determined at association, mobility in the environment and fading will change the SNR of each of the \tagnames over time (see Fig.~\ref{fig:pow_variance}). To address this, each \tagname adjusts its power over time using the signal strength of the query message from the AP, using three different levels. We define the maximum power of the \tagname as 0~dB power gain. First, during association, we consider two cases for the associating device. If it sees a low received signal strength for the AP's query packet, it sets its power gain to the maximum. Otherwise, it sets its gain to the middle level. This gives the higher signal strength backscatter devices leeway to both increase and decrease their power, after association. The AP uses the resulting backscatter signal strengths during association to assign a corresponding cyclic shift. The backscatter devices use the signal strength at association as a baseline and either increase or decrease their power gains for the rest of the concurrent transmissions, i.e., if the signal strength for the AP's query message increases (decreases), the backscatter devices decrease (increase) their power gain. If the \tagname cannot meet its expected SNR requirements given its limited power levels and assigned cyclic shift, it does not join the concurrent transmissions. If this happens more than twice, the backscatter device re-initiates association after which the AP reassigns the cyclic shifts to account for the new significantly different power value (see~\xref{sec:assoc}). 

\begin{figure}[t]
	\centering
	\includegraphics[width=1\columnwidth]{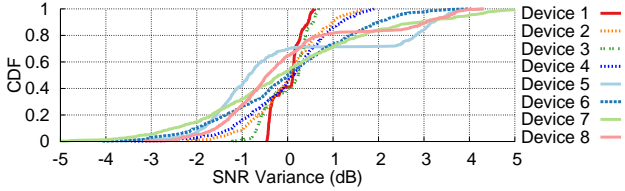}
    \vskip -0.1in
	\caption{{{\bf Backscatter Devices SNR Variance.} CDF of SNR variance of backscatter devices in an office environment, when people were walking around, over 30 mins.}}
	\label{fig:pow_variance}
    \vskip -0.15in
\end{figure}

{The key question however is: how can a low-power backscatter \tagname change its transmission power gain? This is interesting since power adaptation has not been used before {in the network of} backscatter devices. In backscatter, the transmit power gain, $Gain_{power}$, is equal to $\frac{|\Gamma_0-\Gamma_1|^2}{4}$. Here $\Gamma_0$ and $\Gamma_1$ are reflection coefficients for switching between two impedance value, $Z_{0}$ and $Z_{1}$.} Backscatter hardware is designed to maximize the difference between reflection coefficients to maximize their transmission power. This corresponds to $Gain_{power}=0~dB$. One way to achieve this is to switch between extreme impedance values, $Z_{0}=0\Omega$ and $Z_{1}=\infty\Omega$. To achieve power adaptation, in contrast, we pick impedance values that correspond to multiple power settings. In particular, as shown in Fig.~\ref{fig:impedance_plot}, instead of switching from $Z_{0}=0\Omega$, we switch from intermediary impedances and hence achieve lower power gains. Our hardware implementation achieves three power gains of 0~dB, -4~dB and -10~dB to achieve power adaptation. {Note that~\cite{lorabackscatter} uses a similar circuit structure as Fig.~\ref{fig:impedence} to cancel higher order harmonics. We instead design this circuit structure to control the power. }  

{\it Design tradeoff.} Readers might wonder if reducing the power of high SNR devices would decrease the network throughput, since high SNR devices in traditional LoRa backscatter designs can achieve a higher bitrate. In contrast, by reducing their power we are enabling a large number of concurrent transmissions with a fixed  bitrate. Thus, we are encouraging concurrency by reducing the bitrate of high SNR devices.~\xref{sec:deploy} compares the results for NetScatter with one where each backscatter device uses rate adaptation to pick its ideal bitrate, while transmitting alone using LoRa backscatter~\cite{lorabackscatter}.  The results show that the network throughput and latency gains due to large scale concurrency outweigh the reduction in the power for high SNR devices. 


\subsection{\nameDesign Protocol \& Receiver Details}
Pautting it together, the AP transmits an ASK modulated query message which is used to synchronize all the participating concurrent \tagnames. This message conveys information about cyclic shift assignment which are based on the \tagnames' signal strength at the AP.  {The \tagnames measure the query message's signal strength using the envelope detector and use it to fine-tune their transmit power gain.} In the rest of this section, we describe various protocol details required to make our design work in practice. {Note that our focus in the protocol design is about scheduling a set of concurrent transmissions. Typically networks could have more \tagnames than concurrent transmitters supported by our design. Since the AP knows the duty-cycle of each \tagname from the association phase (see~\xref{sec:assoc}), it can i) assign the cyclic shifts and ii) schedule the \tagnames involved in concurrent transmissions.}


\begin{figure}[t]
   	\centering
   	\includegraphics[width=0.9\linewidth]{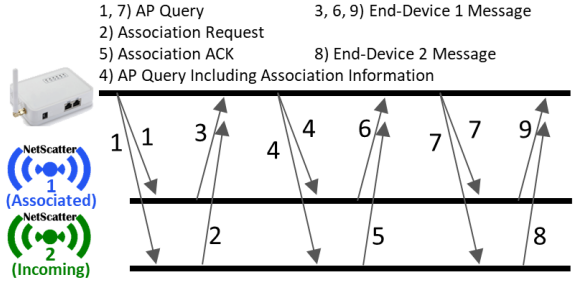}
	\vskip -0.1in
	\caption{{\bf NetScatter Network Association Process.} We show the association process of an incoming NetScatter device (\#2) to the network, while there are existing devices associated with the network (i.e., device \#1).}
   	\label{fig:association}
   	\vskip -0.15in
\end{figure}

\subsubsection{Link-layer Backscatter Packet Structure}
Similar to LoRa, the \tagname packet starts with upchirp and downchirp preambles. They are designed to serve two purposes: i) finding the start of the packet and ii) detecting the transmissions. We emphasize here that the \tagname transmits the same assigned cyclic shift for both upchirps and downchirps in the preamble as well as the payload. The preamble consists of six upchirps followed by $two$ downchirps. This is then followed by the payload and the checksum. We note that in our design, all the \tagnames send their preambles concurrently. This reduces the overhead of transmitting preambles for each \tagname, which in turn increase the end-to-end throughput gain achieved by \nameDesign. The AP uses the above structure to achieve two  goals.

{\vskip 0.05in\noindent {\it i) Finding the exact packet start.} We use the downchirp in the preamble to find the start of the packet transmission. Specifically, we use the middle point between an upchirp and downchirp and switch by six upchirp symbols to the left to find the packet beginning. We suspect that the LoRa preamble has a downchirp for this exact purpose. We note that in our case, since the upchirp and downchirp in the preamble from each of the \tagnames uses the same cyclic shifts, they are symmetric around the middle point and hence the same algorithm for estimating the packet beginning is applied.

\vskip 0.05in\noindent{\it ii) Detecting and decoding each concurrent transmitter.} Now that we found the packet start, we need to find out which transmitters are in the network. To do so, for each preamble symbol, we demodulate it and look at the peaks in FFT domain. If there is an FFT peak in the demodulator output which repeats in all the preamble symbols, we conclude that the \tagname corresponding to that cyclic shift is sending data. After finding current \tagnames in the network, we compute the average power over the six preamble symbols for each \tagname. This average power is used as a threshold to demodulate the payload of each \tagname. In particular, if the power of the \tagname's FFT peak for each payload symbol is more than half this average, we interpret that as $1$ and $0$ otherwise.

\vspace{-10px}
\subsubsection{Network  Association}\label{sec:assoc}
Say the network already has $N$ \tagnames associated to the AP and the $N+1^{th}$ \tagname wants to join the network. A na\"ive approach  is to periodically dedicate time periods for association. This however can lead to high association delays depending on the frequency of the association periods. Our approach instead is to reserve $N_{assoc}$ cyclic shifts and the corresponding FFT bins for association and use the rest for communication. In other words, all the \tagnames transmit at the same time but the ones who want to enter the network transmit with the $N_{assoc}$ association cyclic shifts. 
 
To address the near-far problem, we reserve two cyclic shifts, one in high-SNR and the other one in the low-SNR cyclic shift regions. The incoming \tagname would choose which association region to transmit based on the signal strength of the AP's query message, calculated using the envelope detector. However to account for the hardware delay variations, as before, we skip two cyclic shifts  to ensure that the association packets from the \tagnames can be decoded and won't interfere with communication cyclic shifts. Finally, to support scenarios where more than one \tagname want to associate at the same time, one can use Aloha protocol with binary exponential back-off in the association process. Our deployment does not implement this option and turns ON the backscatter devices one at a time and runs the network only after all the devices are associated.

After the incoming \tagname sends its packet to the AP in association process using the association cyclic shifts, the AP computes its signal strength and decides which cyclic shift and timing schedule it should be assigned to. The AP piggybacks these assignments in its query messages.

\begin{figure}[t]
    \centering
    \includegraphics[width=\linewidth]{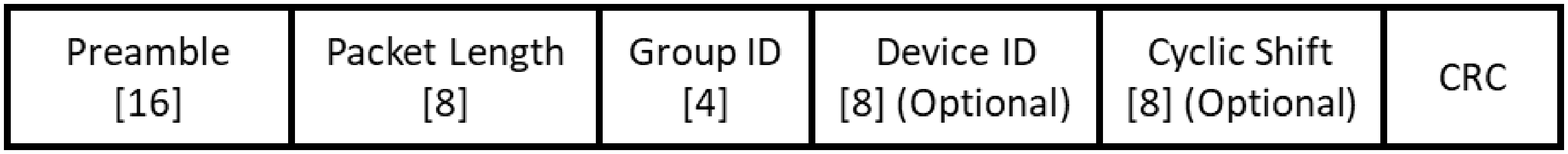}
	\vskip -0.1in
	\caption{{\bf Structure of AP's Query Message.}}
	\label{fig:packet}
    \vskip -0.15in
\end{figure}

{\subsubsection{AP Query Message} \label{dl_offset} Fig.~\ref{fig:packet} shows the ASK-modulated query message that the AP sends. The message has a group ID which identifies the set of 256 devices that should concurrently transmit. In our implementation, since there are only 256 devices, we set this group ID to 0. In a larger network, the AP can assign different sets of devices to different groups depending on their signal strengths, i.e., devices that have a similar signal strength are grouped into the same group to enable concurrent transmissions while further minimizing the near-far problem. 

This is then followed by an optional association response payload that assigns an 8-bit network ID and a 8-bit cyclic shift. Note that prior LoRa backscatter designs are request-response systems that query each backscatter device sequentially and need most of the fields in Fig.~\ref{fig:packet} other than the group ID and cyclic shift assignment. Since these additional 12 bits is transmitted using 160~kbps ASK downlink, the overhead is negligible compared to the 1~kbps backscatter uplink. Finally, we note that if the AP is unable to assign a new device given the existing assignments, the AP updates the cyclic shift assignments for all the devices in the network. It does so by transmitting the identifier for one of the the 256! orderings, which requires $log_2(256!)$ ($\le$1700) bits. This occupies less than 11~ms using our 160~kbps downlink. 

\subsubsection{Network protocol}
Fig.~\ref{fig:association} summarizes our network protocol. First the AP broadcasts its query. \tagName~1, which is already associated to the network receives the query and sends its data  using its assigned cyclic shift after performing any necessary power control. Concurrently, \tagname~2 sends a {\it Association Request}  using one of the $N_{assoc}$ cyclic shifts. The AP receives these two messages and broadcast another query  which includes association information for \tagname~2. Upon receiving this query, \tagName~1 continues to send its data, however, \tagname~2 extract cyclic shift assignment from the query and then transmits {\it Association ACK}  to the AP in the assigned cyclic shift. If AP receives {\it Association ACK}, it adds \tagname~2 to associated devices. Otherwise, it will repeat the association information in the following queries. After association, each \tagname uses its assigned cyclic shift for sending data. }



\section{Evaluation}\label{sec:evaluation}

\begin{figure}[t]
    \centering
    \includegraphics[width=0.85\columnwidth]{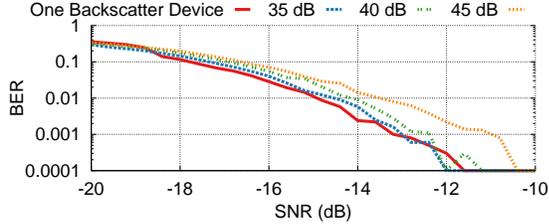}
    \vskip -0.15in
\caption{{\bf Near-Far BER Results.} We show the effect of the second device's power on the first device's BER vs. SNR for different ratios of the second device's to first device's power with power aware cyclic shift assignments.} 
\label{fig:snr_power_diff}
\vskip -0.15in
\end{figure}

\begin{figure}[t!]
    \centering
    \includegraphics[width=0.5\linewidth]{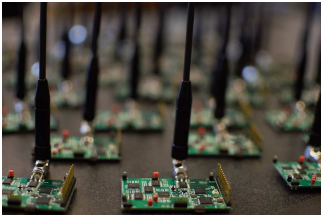}
	\vskip -0.1in
	\caption{{\bf Our Backscatter Devices.} They are arranged closely for this picture. They are spread out across more than ten rooms in our deployment.}
	\label{fig:tags}
    \vskip -0.15in
\end{figure}

\begin{figure}[t]
	\begin{subfigure}{0.49\columnwidth}
    	\centering
    	\includegraphics[width=1.05\linewidth]{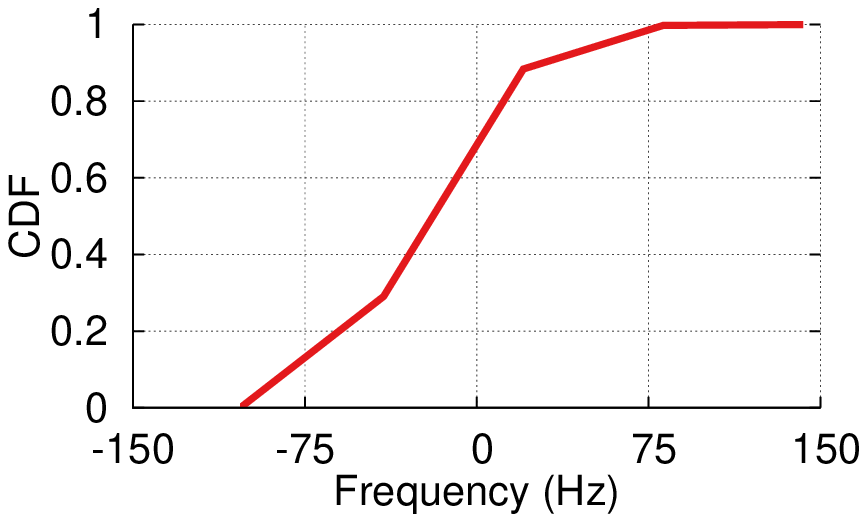}
		\label{fig:FO_passive}
        \vskip -0.2in
		\caption{Frequency Offset} 
		\label{fig:FO}
	\end{subfigure}
    \begin{subfigure}{0.49\columnwidth}
    	\centering
    	\includegraphics[width=1.05\linewidth]{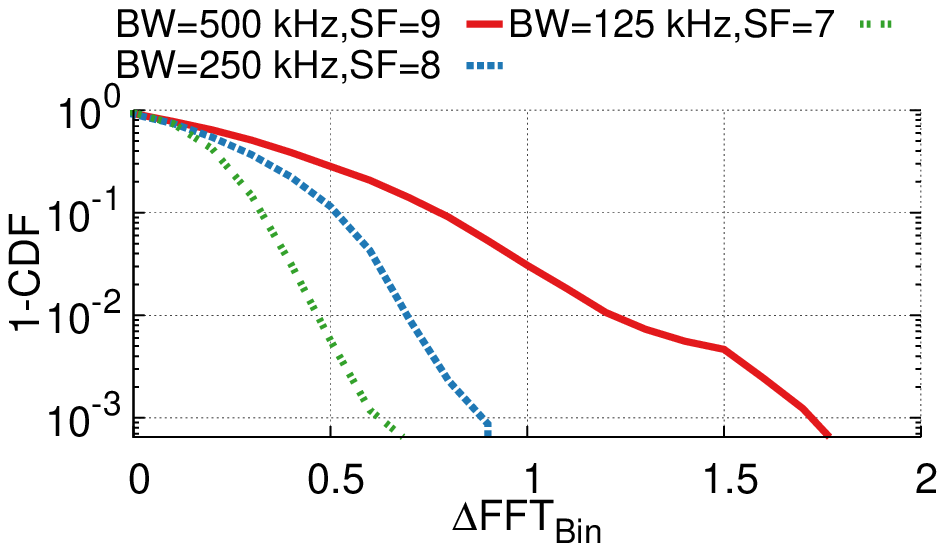}
		\caption{FFT Bin Variation}
		\label{fig:fft_var}
		\vskip -0.1in
	\end{subfigure}
	\vskip -0.1in
	\caption{{{\bf Frequency Offset FFT Bin Variation.} (a) frequency offset of backscatter devices, and (b) effect of residual time and frequency offset for different configurations.}}
	\label{fig:FO_fft_var}
	\vskip -0.15in
\end{figure}




{
\subsection{Hardware implementation}\label{sec:implementation}
\noindent{\bf COTS Implementation.} Our COTS hardware shown in Fig.~\ref{fig:tags} consists of RF section and baseband section, both implemented on a four layers FR4 PCB. On RF receive side, we implemented envelope detector similar to~\cite{nsdi16} but at 900~MHz and it has a sensitivity of -49~dBm  to receive downlink query messages from AP.\footnote{Note that since ASK-modulated AP query received by backscatter \tagname experiences one-way path loss, its required sensitivity is only $-44~dBm$ in contrast to the -120~dBm sensitvity for the backscatter signals.} RF transmit side consists of five ADG904~\cite{adg904} switches cascaded in three levels to build an impedance switch network for backscatter, power gain control and also switching between transmit and receive modes. Our backscatter \tagname uses a 2~dBi whip antenna to transmit packets and receive query messages in the 900~MHz ISM band.}
{The baseband side is implemented using an IGLOO nano AGLN250 FPGA~\cite{igloo} and an MSP430FR5969~\cite{msp430}. We generate CSS packets on the FPGA and output real and imaginary components of the square wave signal to the backscatter switch network. The envelope detector is controlled by the MCU. Downlink receiver algorithm is implemented on MCU. To be resilient to self-interference caused by the AP's single-tone, the baseband at the backscatter device shifts the AP's signal by 3~MHz. Note that the COTS implementation is for prototyping and proof-of-concept; an ASIC is typically required to achieve the orders of magnitude power benefits of backscatter communication.}

{\vskip 0.05in\noindent{\bf IC Simulation.} We designed and simulated an IC for our backscatter device using TSMC 65nm LP process. It consists of four blocks with total power consumption of 45.2~$\mu W$: i) An envelope detector that demodulates the APs ASK query messages and consumes less than 1~$\mu W$. ii) Baseband processor for processing and extracting AP data from envelope detector, interfacing with sensors and sending the chirp specifications and sequence of data to chirp generator is handled by this block consuming 5.7~$\mu W$ of power. iii) A chirp generator that takes SF, BW, cyclic shift assignment and data sequence from the baseband processor to generate the sequence of ON-OFF keying chirps. We used Verilog code to describe the baseband signal's phase behavior and generate assigned cyclic shift with required frequency offset. We used Synthesis, Auto-Place and Route (SAPR) to simulate Verliog code on chip. The power consumption of this block is 36~$\mu W$.} iv) 
A Switch network which is composed of three resistors that are connected to NMOS switches to generate backscatter signal with three power gain levels. Note that since these resistors and NMOS switches consume minimal area, more power gain levels can be added at almost no cost. The power consumption of the switch network is 2.5~$\mu W$ with 3~MHz frequency offset.}

{\vskip 0.05in\noindent{\bf Reader Implementation.} We implement the reader on the X-300 USRP software-defined radio platform by Ettus Research~\cite{usrpX300}. We use a mono-static radar configuration with two co-located antennas separated by 3~feet. The transmit antenna is connected to a UBX-40 daughterboard, which transmits the query message and the single-tone signal. The USRP output power is set at 0~dBm and we use an RF5110 RF power amplifier~\cite{amplifier} to amplify the transmit signal to 30~dBm. The receiver antenna is connected to another UBX-40 daughterboard, which down-converts the NetScatter packets to baseband signal and samples it at 4~Msps.}



\subsection{Frequency and Timing Mismatch}\label{sec:tofo_eval}

{\it Measurements 1: Hardware frequency variations.} We measure the frequency offsets of our hardware by recording thousand packets for each \tagname. Using the method described in \xref{dl_offset}, we compute the frequency offset for the 256 backscatter \tagnames in our network deployment which we show in Fig.~\ref{fig:FO}. { The variations of backscatter \tagnames are less than 150~Hz which is nearly 0.15th of one FFT bin when $BW=500kHz$ and $SF=9$. Therefore, our system is not affected by frequency variation of different \tagnames.}

\begin{figure}[t]
	\begin{subfigure}{0.49\columnwidth}
		\centering
		\includegraphics[width=1.05\linewidth]{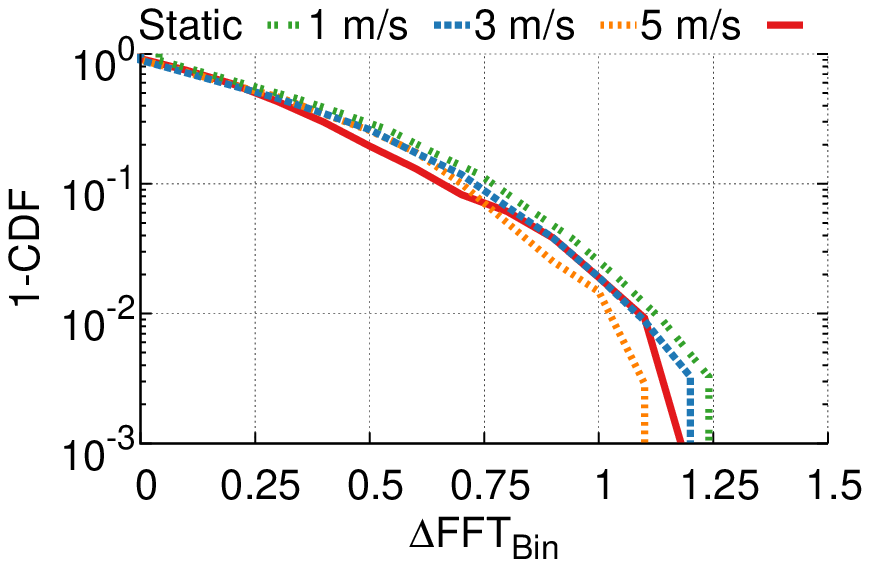}
    	\vskip -0.1in
		\caption{Doppler Effect}
		\label{fig:doppler}
	\end{subfigure}
    \begin{subfigure}{0.49\columnwidth}
		\includegraphics[width=1.05\linewidth]{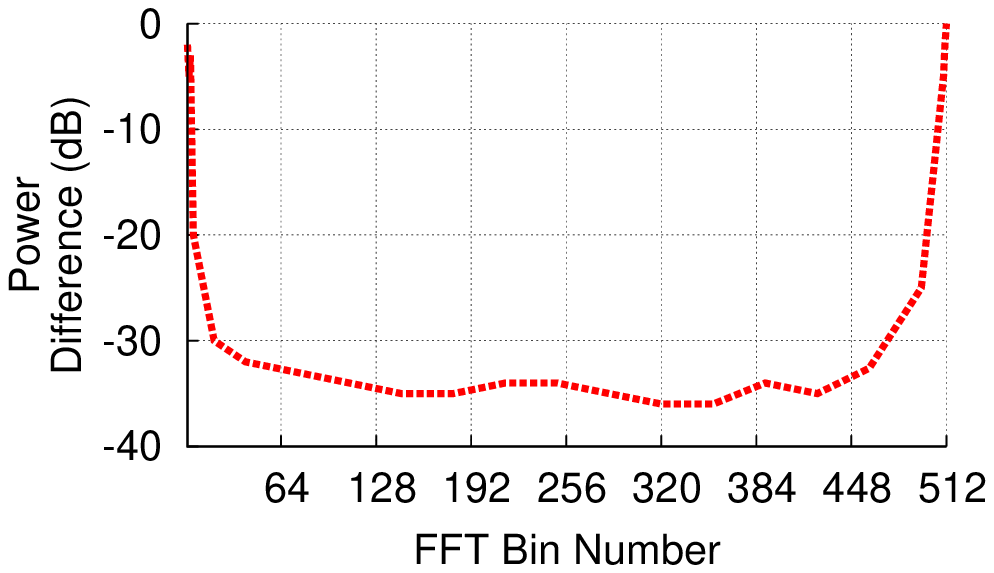}
		\caption{Power Dynamic Range}
		\label{fig:pow_dyn}
	\end{subfigure}
    \vskip -0.1in
	\caption{{{\bf Doppler Effect and Power Dynamic Range Evaluation.} We evaluate (a) Doppler effect, and (b) we show power difference between two concurrent transmissions at different locations of FFT domain. One transmission is fixed and the other is sweeping across different chirp symbols.}}
	\label{fig:doppler_pow_dyn}
    \vskip -0.15in
\end{figure}

{\it Measurements 2: Timing offsets.}  Next, we characterize how the timing offsets affects $\Delta FFT_{bin}$. This helps us  understand how many empty cyclic shifts, $SKIP-1$, we need to put for each occupied cyclic shift. To do this, we setup a wireless experiment sending query messages from the AP and receiving transmissions from the backscatter \tagnames deployed in our system. By decoding these transmissions and comparing the received cyclic shifts with what we have programmed the \tagnames to send, we can find the $\Delta FFT_{bin}$ for each \tagname; this measurement is a combination of  both timing and the small frequency variations on the hardware.

Fig.~\ref{fig:fft_var} shows residual $\Delta FFT_{bin}$ for backscatter \tagnames. {The plots show that the $\Delta FFT_{bin}$ is considerable. This is because in backscatter \tagnames, the energy detector receives the amplitude modulated query message and sends interrupt to initiate backscatter transmission. Both these steps add to the timing variations. Specifically, the hardware delay variation comes from variation in receiving query message and initiating the transmission on FPGA which can vary from packet to packet.} In our deployment in~\xref{sec:deploy} with backscatter \tagnames, we use BW=500~kHz, SF=9 and leave one FFT bin between occupied cyclic shifts ($SKIP=2$). This translates to supporting 256 \tagnames with an aggregate throughput of around 250~kbps and bitrate per tag of around 1~kbps.


{\it Measurements 3: Doppler effects.}  Other than hardware frequency offsets, Doppler effect can cause changes in frequency as well. However, the effect of it will be much less than 1 FFT bin, $\frac{BW}{2^{SF}}$, for most cases. As an example, assume a backscatter device is moving with a speed of 10~m/s. Considering the carrier frequency is 900~MHz, the doppler effect induced frequency change would be 30~Hz which is much less than 1~kHz, the FFT bin frequency, assuming BW=500~kHz and SF=9. 
To confirm this, we run various mobility experiments where a subject holds a backscatter \tagname and moves with different average speeds which we measure using an accelerometer. We receive transmissions from the \tagname and compute the $\Delta FFT_{bin}$ for different motion scenarios. Fig.~\ref{fig:doppler} shows $\Delta  FFT_{bin}$ for various speeds, which confirms that these speeds do not have an effect on $\Delta FFT_{bin}$.

\begin{figure}[t!]
	\begin{subfigure}{0.325\columnwidth}
    	\centering
    	\includegraphics[width=\linewidth]{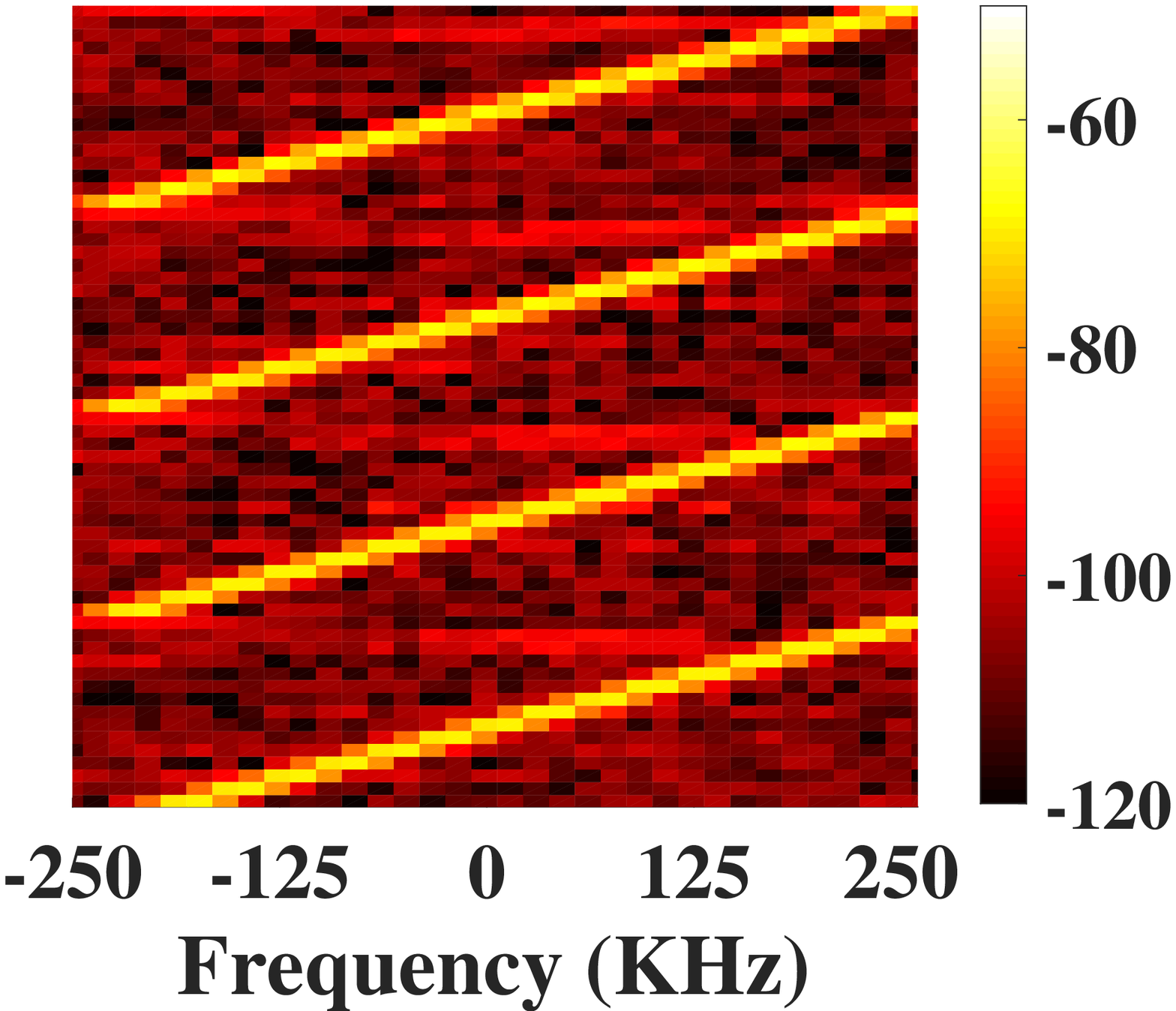}
		\label{fig:power_spec_high}
        \vskip -0.2in
        \caption{High}
	\end{subfigure}
    \begin{subfigure}{0.325\columnwidth}
    	\centering
    	\includegraphics[width=\linewidth]{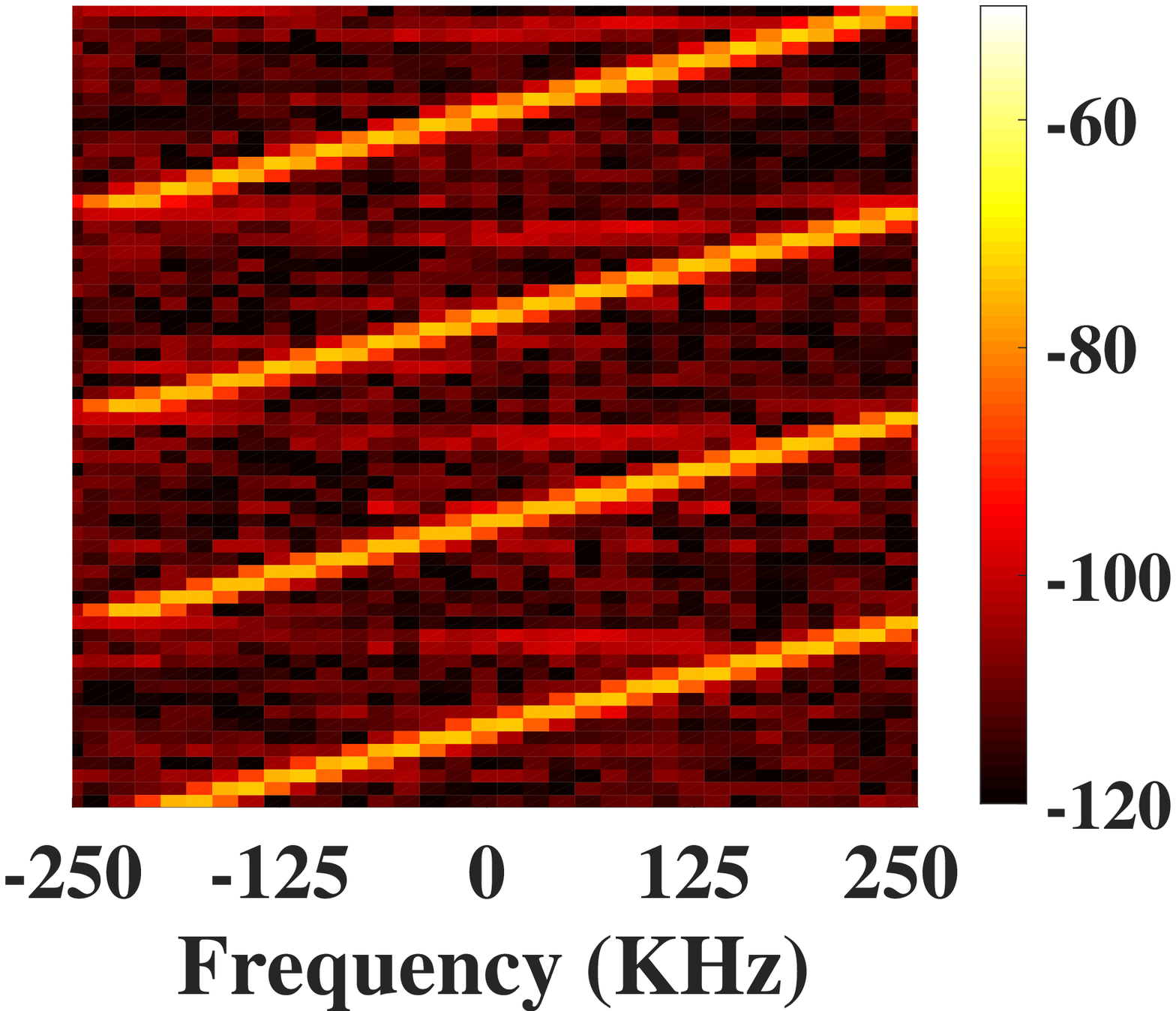}
		\label{fig:power_spec_med}
        \vskip -0.2in
        \caption{Medium}
	\end{subfigure}
    \begin{subfigure}{0.325\columnwidth}
    	\centering
    	\includegraphics[width=\linewidth]{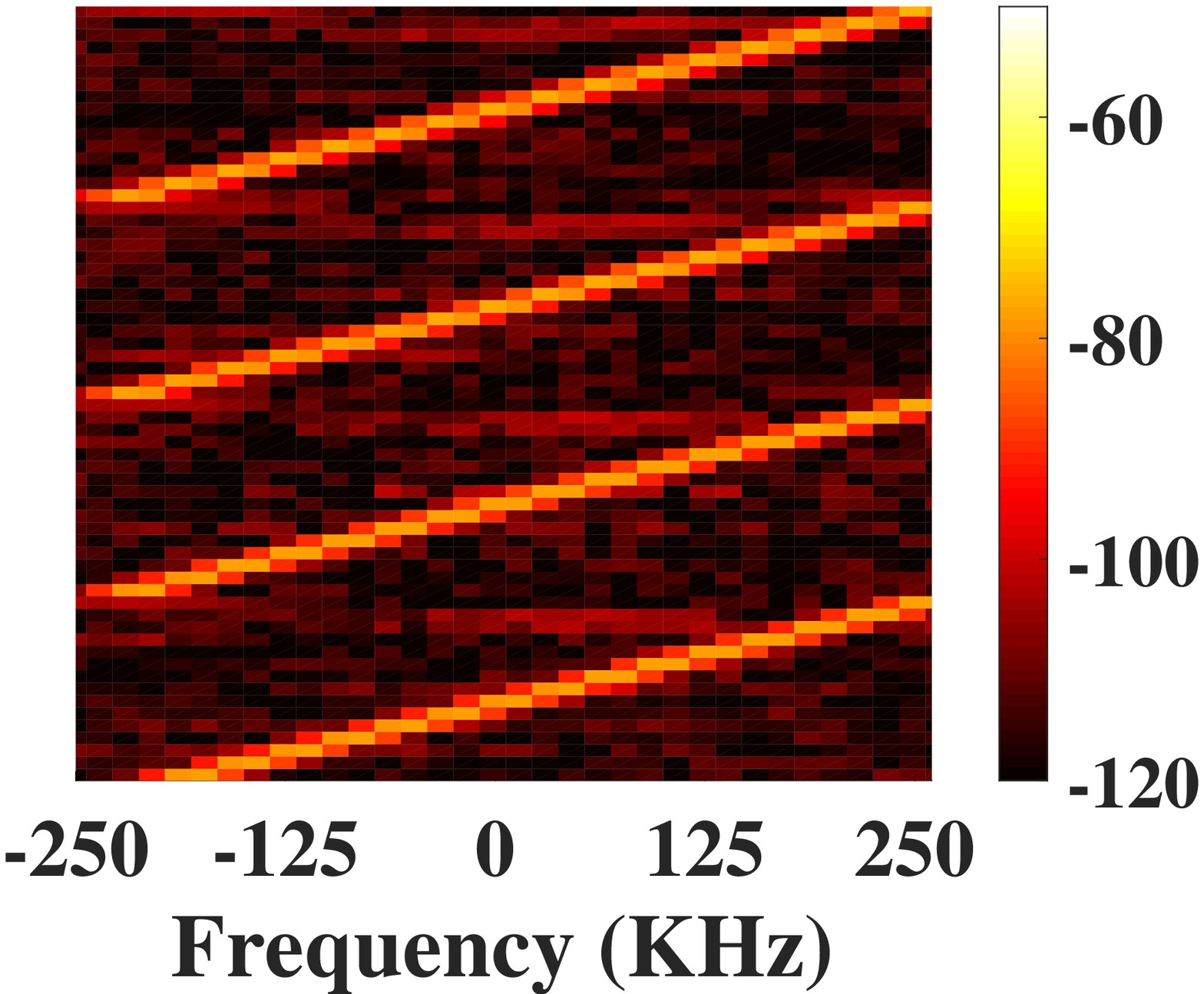}
		\label{fig:power_spec_low}
        \vskip -0.2in
        \caption{Low}
	\end{subfigure}
\vskip -0.1in
\caption{{\bf Spectrogram of Backscattered Signal at the Different Power Levels.}}
\label{fig:power_spec}
\vskip -0.15in
\end{figure}

\subsection{Near-Far Problem}\label{nearfar_eval} 

\vskip 0.01in
{\it Measurements 1: Power-aware cyclic shift assignment.}  As mentioned in ~\xref{sec:nearfar}, we assign cyclic shifts to \tagnames depending on their signal strength values. To evaluate the effectiveness of this technique, we run experiments with two \tagnames where one of them transmits at a high power (equivalent to being near the AP) with a cyclic shift corresponding to the beginning of the FFT spectrum. Then, we sweep the cyclic shift of the second \tagname from small FFT bin difference cyclic shifts to high FFT bin difference ones. At each of the cyclic shifts, we decrease the power of the second device using an attenuator up to when it has packet error rates less than one percent.  Fig.~\ref{fig:pow_dyn} shows the maximum power difference that can be tolerated  between these two devices versus the assigned FFT bin difference. As can be seen, as we go further in FFT bin difference, we can tolerate more power difference between the two \tagnames. Note that, because of aliasing Fig.~\ref{fig:pow_dyn} is symmetric around the center. The maximum happens in middle and is equal to 35~dB. This is the dynamic range that our system can support in practice. We also note that when the second device is assigned to an FFT bin 2 cyclic shifts away from the first device, it can be up to 5~dB below the latter and still correctly decoded. This means there is an in-built 5~dB dynamic range resilience to channel variations between devices that have close cyclic shifts.

{\it Measurements 2: Self-aware power-adjustment.} The second method to address the near-far problem and also increase the dynamic-range is power adjustments at the \tagnames using the signal strength of the AP's query message. To evaluate this, we first measure how well we can adjust power on the \tagnames. We evaluate its efficacy in practical deployments. We use three different backscatter impedance values to be able to transmit packets in three different power gains. Fig.~\ref{fig:power_spec} shows the spectrum of backscattered signal at different power levels. These plots show that the hardware creates spectrum that is clean and does not introduce noticeable non-linearities into the backscattered signal. Furthermore, we can achieve three different power levels: 0, -4, and -10~dB.


\subsection{Network Deployment}\label{sec:deploy}
We evaluate three key network parameters: 
\squishlist
\item {\bf Network PHY bit-rate.} This is the bitrate achieved across all the devices during the payload part of the packet.
\item {\bf Link-layer data rate.} This is the data rate achieved in the network which is defined as the data rate for sending useful payload bits, after considering overheads including the AP's query message and the preamble of the packet transmission.
\item {\bf Network latency.} This is the latency to get the payload bits from all the backscatter devices in the network. 
\squishend

\begin{figure}[t]
    \centering
    \includegraphics[width=1\linewidth]{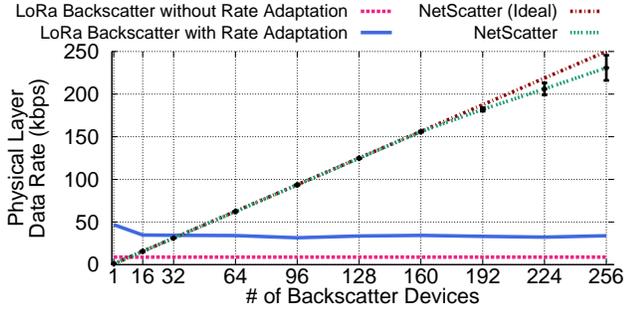}
    \caption{{\bf Network Physical Rate.} We evaluate \nameDesign network physical rate and compare it with other schemes.}
	\vskip -0.1in
	\label{fig:passive_phy}
\end{figure}

We compare three schemes: i) LoRa backscatter~\cite{lorabackscatter} where all devices use a fixed bitrate {of 8.7~kbps}, ii) LoRa backscatter with rate adaptation where each device uses the best bitrate given its channel conditions and iii) \nameDesign.   Note that the authors of~\cite{lorabackscatter} did not publicly release the code and so, we replicate the implementation adding the missing details and using $BW=500~kHz$ and $SF=9$. We also note that~\cite{lorabackscatter} is not designed to work with more than one to two users. Here, we use query-response design with scheduling when there are more users where the AP queries each \tagname. While LoRa backscatter does not support rate adaptation, we wanted to compare with an ideal approach that maximizes the bitrate of each \tagname by picking the optimal $SF$ and $BW$. To do so, we measure the signal strength from each of the backscatter \tagnames and compute the bitrate using the SNR table in~\cite{lora1276}; this is the ideal performance a single-user LoRa backscatter design achieves with rate adaptation.

{\bf Network PHY bit-rate.} We set each \tagname bit-rate to  976~bps, $BW_{agg}=500~kHz$, $SF=9$ and a payload size of five bytes. We deploy 256 backscatter \tagnames across the floor of an office building with more than ten rooms. Fig.~\ref{fig:floormap} shows our deployment in an office.
Fig.~\ref{fig:passive_phy} shows the results of network physical rate for our backscatter network deployment. The plot highlights the following key observations.
\squishlist
\item The network data rate scales with the number of concurrent backscatter devices. When the number of concurrent devices is less than 128, the variance in the throughput is small. This is because in these scenarios  effectively the backscatter devices are separated from each other by more than 2 cyclic shifts (SKIP $\ge 3$). As a result, the devices do not interfere with each other and hence can concurrently operate. As we increase the concurrent devices to 256, we are pushing the system to its theoretical limit (with SKIP = 2) and thus, we see larger variances in the network data rate.
\item With 256 backscatter \tagnames, \nameDesign increases the PHY bit-rate by 6.8x and 26.2x over LoRa backscatter with and without rate adaptation. The gains are lower with the ideal rate adaptation since with rate adaptation high-SNR devices could pick the maximum LoRa bitrate of 32~Kbps.  
\squishend

\begin{figure}[t]
    \centering
    \includegraphics[width=1\linewidth]{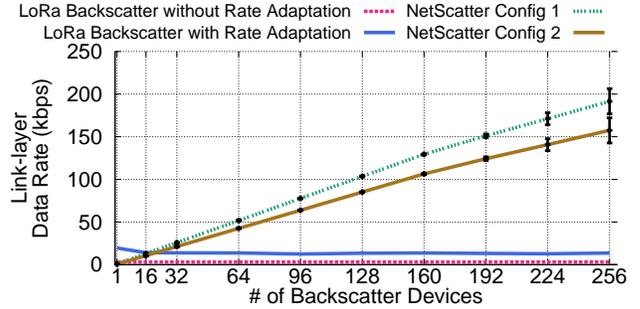}
    \vskip -0.1in
    \caption{{\bf Link-layer Data Rate.} We evaluate link-layer data rate for \nameDesign and compare it with other schemes.}
    \vskip -0.1in
	\label{fig:passive_link}
\end{figure}

{\bf Link-layer data rate.} While the above plots measure the data rate improvements for the message payload, it does not account for the end-to-end overheads including preambles and the AP's query message to coordinate the concurrent transmissions. To see the effect of the AP query packet overhead for \nameDesign, we consider two configurations.
\squishlist
\item {\it NetScatter Config 1.} In this scenario the cyclic shifts are all assigned during the association phase and the AP query packet coordinating the concurrent transmissions is 32~bits long without the optional fields in Fig.~\ref{fig:packet}. 
\item {\it NetScatter Config 2.} In this scenario, the AP query packet contain cyclic shift assignments for all the devices in the network and has a length of 1760~bits. 
\squishend
The above two configurations represent the two extremes of our deployment.  {We set the backscatter payload and CRC to 40~bits and use the total 8 upchirps and downchirps for preamble.  For LoRa backscatter which queries each individual device sequentially, the AP query is 28 bits long.}

{ Fig.~\ref{fig:passive_link} shows that the gains at the link-layer are higher for \nameDesign  over LoRa backscatter without and with rate adaptation by 61.9x (50.9x) and 14.1x (11.6x) respectively for config\#1 (\#2). This is because, in \nameDesign, the added overhead of \tagnames' preambles happen once and at the same time for all \tagnames. But the other schemes need to do TDMA which means that sending preamble will not happen concurrently for all \tagnames and these have to be sent individually for each backscatter device since in traditional designs the AP querying each of them sequentially. Further, in LoRa backscatter which queries sequentially, the AP query message is transmitted once for each device in the network versus being transmitted once for all the devices in our design. Finally, since the downlink uses ASK at 160~kbps, the overhead of transmitting 1760~bits in config\#2, while reducing the link-layer data rate over config\#1, is still {low} because the backscatter links can only achieve a much lower bitrate.}

{\bf Network latency.}
{Finally, Fig.~\ref{fig:passive_latency} shows that \nameDesign has a latency reduction of 67.0x (55.1x) and 15.3x (12.6x) over prior LoRa backscatter without and with rate adaptation respectively in network config\#1 (\#2). This is the key advantage of using concurrent transmissions in low-power backscatter networks. It is noteworthy that since the downlink AP query bit-rate is 160~kbps, AP query duration is negligible compared to duration of backscatter devices' preamble for prior backscatter methods and also for config\#1. For config\#2, the AP query duration is significantly higher than the config\#1. However, the total duration is still dominated by the backscatter payload + CRC and preamble. As a result, AP query is not the dominant factor in link-layer latency.}

\begin{figure}[t]
    \centering
    \includegraphics[width=1\linewidth]{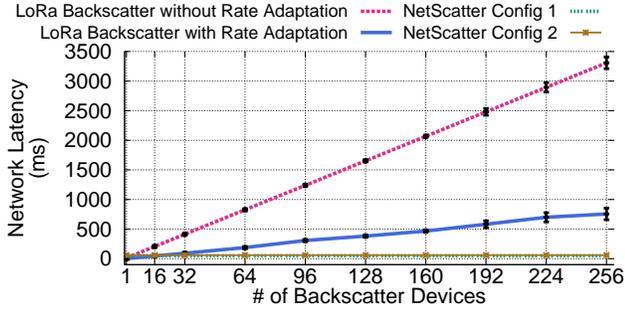}
    \vskip -0.1in
	\caption{{\bf Network Latency.} We evaluate the latency of \nameDesign and compare it with other schemes.}
    \vskip -0.15in
	\label{fig:passive_latency}
\end{figure}


\section{Related Work}\label{sec:related}

Recent systems use backscatter with Wi-Fi  signals~\cite{sachin1,fsbackscatter,nsdi16}, have a receiver sensitivity of only -90~dBm and hence have a limited range and cannot work across rooms unless the RF source is placed close to the backscatter tag~\cite{conext17-sachin,nsdi16}.  LoRa backscatter~\cite{lorabackscatter} can achieve long ranges by generating LoRa-compliant packets at the backscatter device. pLoRa~\cite{plorasigcomm2018} backscatters ambient LoRa signals in the environment in contrast to the single tone used as the RF source in NetScatter as well as~\cite{lorabackscatter}. We note that all SemTech LoRa chipsets have the capability in software to transmit single tone signals. All these prior long range systems are evaluated in a network of only 1--2 devices and propose to use time-division to support multiple backscatter devices. In contrast, our design enables large-scale concurrent transmissions and can achieve much higher link-layer data rates as well as lower latencies. We also note that these long range backscatter systems~\cite{lorabackscatter,plorasigcomm2018} claim a kilometer range in outdoor scenarios such as open fields. This however requires placing the RF source close to the backscatter devices. In indoor environments where the signal propagates through walls and the RF source is not placed close to the backscatter devices,  our network operational range across ten different rooms is consistent with these prior work. Finally, we note that while prior work~\cite{lorabackscatter,plorasigcomm2018} decodes the backscatter signal on Semtech LoRa chipsets, our distributed CSS protocol is decoded on a software radio. We however note that SemTech LoRa SX1257~\cite{sx1257} chipsets provide I-Q samples and hence our approach could also be implemented on these off-the-shelf chipsets {together with a low power FPGA for baseband processing}; this however is not in the scope of this paper.


Finally, recent work on decoding concurrent transmissions from RFID tags, does not achieve the long range operations and below-noise operations of CSS based systems. Buzz~\cite{buzz}, LF-Backscatter~\cite{lfbackscatter}, and others~\cite{bst,rfidgroup,fliptracer} leverage the differences in the time domain signal transitions and changes in the constellation diagram to decode multiple RFIDs.  However, the number of concurrent transmissions in the above designs is limited  --- the latest in this line of work, Fliptracer~\cite{fliptracer}, can reliably decode up to five concurrent RFID tags. Further, these systems were tested with ranges of 0.5 to 6 feet~\cite{buzz,lfbackscatter,fliptracer} and in the same room. Finally, receiver sensitivity of even battery-powered backscatter tags for RFID EPC-GEN2 readers is around -85~dBm. So it cannot support the long ranges and whole-home deployments that CSS modulation based backscatter achieves. 



\section{Conclusion}\label{sec:conclusion}
{We present a new wireless protocol for backscatter networks that scales to hundreds of concurrent transmissions. To this end, we introduce, distributed chirp spread spectrum coding, which uses a combination of chirp spread spectrum (CSS) modulation and ON-OFF keying. Further, we  address practical issues including near-far problem and timing and frequency synchronization. Finally, we deploy our system in an indoor environment with 256 concurrent devices to demonstrate its throughput and latency performance.}





{\footnotesize 
\balance
\bibliographystyle{abbrv}
\bibliography{ourbib}

\begin{thebibliography}{10}

\bibitem{igloo}
Igloo nano fpga datasheet by, 2015.
\newblock
  \url{https://www.microsemi.com/document-portal/doc_download/130695-igloo-nano-low-power-flash-fpgas-datasheet}.

\bibitem{crystalPPM}
Xrcha-f-a series, 2015.
\newblock
  \url{https://www.murata.com/~/media/webrenewal/products/timingdevice/crystalu/flyers/vppt-hcrj078-d.ashx?la=en-us}.

\bibitem{adg904}
Adg904 datasheet by analog devices, 2016.
\newblock
  \url{http://www.analog.com/media/en/technical-documentation/data-sheets/ADG904.pdf}.

\bibitem{lora1276}
Sx1276 datasheet by semtech, 2016.
\newblock \url{https://www.semtech.com/uploads/documents/sx1276.pdf}.

\bibitem{msp430}
Msp430fr5969 datasheet by ti, 2017.
\newblock \url{http://www.ti.com/lit/ds/symlink/msp430fr5969.pdf}.

\bibitem{amplifier}
Rf5110 rf power amplifier, 2018.
\newblock
  \url{http://www.rfmd.com/store/downloads/dl/file/id/30508/5110g_product_data_sheet.pdf}.

\bibitem{sx1257}
Sx1257 datasheet by semtech, 2018.
\newblock \url{https://www.semtech.com/uploads/documents/DS_SX1257_V1.2.pdf}.

\bibitem{usrpX300}
Usrp x-300, 2018.
\newblock \url{https://www.ettus.com/product/ details/X300-KIT}.

\bibitem{chirppaper}
A.~Berni and W.~Gregg.
\newblock On the utility of chirp modulation for digital signaling.
\newblock {\em IEEE Transactions on Communications}, 21(6):748--751, Jun 1973.

\bibitem{sachin1}
D.~Bharadia, K.~R. Joshi, M.~Kotaru, and S.~Katti.
\newblock Backfi: High throughput wifi backscatter.
\newblock SIGCOMM '15.

\bibitem{devasirvatham1984time}
D.~M. Devasirvatham.
\newblock Time delay spread measurements of wideband radio signals within a
  building.
\newblock {\em Electronics Letters}, 20(23):950--951, 1984.

\bibitem{lorasigcomm17}
R.~Eletreby, D.~Zhang, S.~Kumar, and O.~Ya\u{g}an.
\newblock Empowering low-power wide area networks in urban settings.
\newblock SIGCOMM '17.

\bibitem{lfbackscatter}
P.~Hu, P.~Zhang, and D.~Ganesan.
\newblock Laissez-faire: Fully asymmetric backscatter communication.
\newblock In {\em Proceedings of the 2015 ACM Conference on Special Interest
  Group on Data Communication}, SIGCOMM '15.

\bibitem{bst}
P.~Hu, P.~Zhang, and D.~Ganesan.
\newblock Leveraging interleaved signal edges for concurrent backscatter.
\newblock In {\em Proceedings of the 1st ACM workshop on Hot topics in
  wireless}, pages 13--18. ACM, 2014.

\bibitem{interscatter}
V.~Iyer, V.~Talla, B.~Kellogg, S.~Gollakota, and J.~Smith.
\newblock Inter-technology backscatter: Towards internet connectivity for
  implanted devices.
\newblock In {\em Proceedings of the 2016 ACM SIGCOMM Conference}.

\bibitem{fliptracer}
M.~Jin, Y.~He, X.~Meng, Y.~Zheng, D.~Fang, and X.~Chen.
\newblock Fliptracer: Practical parallel decoding for backscatter
  communication.
\newblock In {\em Proceedings of the 23rd Annual International Conference on
  Mobile Computing and Networking}, MobiCom '17.

\bibitem{wifibackscatter}
B.~Kellogg, A.~Parks, S.~Gollakota, J.~R. Smith, and D.~Wetherall.
\newblock Wi-fi backscatter: Internet connectivity for rf-powered devices.
\newblock In {\em Proceedings of the 2014 ACM Conference on SIGCOMM}.

\bibitem{nsdi16}
B.~Kellogg, V.~Talla, S.~Gollakota, and J.~R. Smith.
\newblock Passive wi-fi: Bringing low power to wi-fi transmissions.
\newblock In {\em 13th {USENIX} Symposium on Networked Systems Design and
  Implementation ({NSDI} 16)}.

\bibitem{conext17-sachin}
M.~Kotaru, P.~Zhang, and S.~Katti.
\newblock Localizing low-power backscatter tags using commodity wifi.
\newblock In {\em CoNext'17}.

\bibitem{abc}
V.~Liu, A.~Parks, V.~Talla, S.~Gollakota, D.~Wetherall, and J.~R. Smith.
\newblock Ambient backscatter: Wireless communication out of thin air.
\newblock SIGCOMM '13.

\bibitem{rfidgroup}
J.~Ou, M.~Li, and Y.~Zheng.
\newblock Come and be served: Parallel decoding for cots rfid tags.
\newblock In {\em Proceedings of the 21st Annual International Conference on
  Mobile Computing and Networking}, pages 500--511. ACM, 2015.

\bibitem{plorasigcomm2018}
Y.~Peng, L.~Shangguan, Y.~Hu, Y.~Qian, X.~Lin, X.~Chen, D.~Fang, and
  K.~Jamieson.
\newblock {PLoRa}: Passive long-range data networks from ambient lora
  transmissions.
\newblock SIGCOMM '18.

\bibitem{saleh1987statistical}
A.~A. Saleh and R.~Valenzuela.
\newblock A statistical model for indoor multipath propagation.
\newblock {\em IEEE Journal on selected areas in communications},
  5(2):128--137, 1987.

\bibitem{sornin2017signal}
N.~Sornin and L.~Champion.
\newblock Signal concentrator device, Oct.~17 2017.
\newblock US Patent 9,794,095.

\bibitem{lorabackscatter}
V.~Talla, M.~Hessar, B.~Kellogg, A.~Najafi, J.~R. Smith, and S.~Gollakota.
\newblock Lora backscatter: Enabling the vision of ubiquitous connectivity.
\newblock {\em Proc. ACM Interact. Mob. Wearable Ubiquitous Technol.}, 2017.

\bibitem{davidtse}
D.~Tse and P.~Viswanath.
\newblock {\em Fundamentals of wireless communication}.
\newblock Cambridge university press, 2005.

\bibitem{lorea-europe}
A.~Varshney, O.~Harms, C.~M. P{\'{e}}rez{-}Penichet, C.~Rohner, F.~Hermans, and
  T.~Voigt.
\newblock Lorea: A backscaer architecture that achieves a long communication
  range.
\newblock Sensys'17.

\bibitem{buzz}
J.~Wang, H.~Hassanieh, D.~Katabi, and P.~Indyk.
\newblock Efficient and reliable low-power backscatter networks.
\newblock In {\em Proceedings of the ACM SIGCOMM 2012 conference on
  Applications, technologies, architectures, and protocols for computer
  communication}, pages 61--72. ACM, 2012.

\bibitem{fsbackscatter}
P.~ZHANG, M.~Rostami, P.~Hu, and D.~Ganesan.
\newblock Enabling practical backscatter communication for on-body sensors.
\newblock In {\em Proceedings of the 2016 ACM SIGCOMM Conference}.

\end{thebibliography}

}


\end{document}